\documentclass[12pt, centerh1]{article}
\usepackage{amsthm,amsmath,natbib,algorithm}
\RequirePackage[colorlinks,citecolor=blue,urlcolor=blue]{hyperref}
\usepackage{natbib}
\usepackage{graphicx}
 \usepackage{multirow}
 \usepackage{caption}
 \usepackage{color}
\usepackage{subfigure}
\usepackage{float}
\usepackage{algorithmic}
\newcommand{\vecx}{\mathbf{x}}
\newcommand{\vecX}{\mathbf{X}}
\newcommand{\vecR}{\mathbf{R}}
\newcommand{\vecC}{\mathbf{C}}
\newcommand{\vecq}{\mathbf{q}}

\title{Longitudinal Data Clustering with a Copula Kernel Mixture Model}
\author{Xi Zhang\textsuperscript{1}, Orla A. Murphy\textsuperscript{2}*, \\ and Paul D. McNicholas\textsuperscript{1}}
\date{\small \textsuperscript{1}Department of Mathematics and Statistics, McMaster University, Ontario, Canada\\
\textsuperscript{2}Department of Mathematics and Statistics, Dalhousie University, Nova Scotia, Canada\\
*Corresponding author. E-mail: orla.murphy@dal.ca}

\begin{document}
\maketitle

\begin{abstract}
Many common clustering methods cannot be used for clustering multivariate longitudinal data in cases where variables exhibit high autocorrelations. In this article, a copula kernel mixture model (CKMM) is proposed for clustering data of this type. The CKMM is a finite mixture model which decomposes each mixture component's joint density function into its copula and marginal distribution functions. In this decomposition, the Gaussian copula is used due to its mathematical tractability and Gaussian kernel functions are used to estimate the marginal distributions.  A generalized expectation-maximization algorithm is used to estimate the model parameters. The performance of the proposed model is assessed in a simulation study and on two real datasets. The proposed model is shown to have effective performance in comparison to standard methods, such as $K$-means with dynamic time warping clustering and latent growth models. 

\end{abstract}\vspace{0.5cm}

\newpage
\section{Introduction}
Longitudinal data, also called panel data, are multi-variable observations recorded for multiple subjects over time. This kind of data exists widely in various fields, such as economics, social sciences, finance and medicine. Due to the structure of longitudinal data, it can provide more information for clustering than either cross-sectional data or univariate time series alone. 

Although distance-based methods can be used to cluster longitudinal data, common distance functions, such as Euclidean and Minkowski distances, do not consider serial dependence in a time series. \cite{berndt1994using} proposed a specific method called dynamic time warping (DTW) to measure the similarity of two sequences with different frequencies. \cite{Genolini15} developed the $\mathtt{Kml}$ {\sf R} package which includes a $K$-means clustering algorithm using DTW for longitudinal data clustering. As alternative to computing distances between sequences, the distances of coefficients from longitudinal data regression models may also be used \citep{den2020comparison}. However, ignoring correlations across time is a limitation in distance-based clustering.

Model-based clustering, which uses finite mixture models, overcomes this drawback. A random sample $\vecX_{1},\dots,\vecX_{N}$ arises from a finite mixture model (FMM) with $G$ components if the  probability density function of each $\vecX_{n}$ can be written as
\begin{equation}
f(\vecx_{n}\mid\boldsymbol{\Theta})=\sum_{g=1}^{G}\pi_{g}f(\vecx_{n}\mid \boldsymbol{\theta}_{g}),
\label{eq:finitemix}
\end{equation}
where 
$f(\cdot\mid \boldsymbol{\theta}_{g})$ is the $g$th component density function with class-specific parameter vector $\boldsymbol{\theta}_{g}$, $\pi_{g}>0$ are the mixing proportions such that $\sum_{g=1}^{G}\pi_{g}=1$, and $\boldsymbol{\Theta}=\{\pi_{1},\dots,\pi_{G},\boldsymbol{\theta}_1,\dots, \boldsymbol{\theta}_G\}$ is the collection of all model parameters. The finite mixture model assumes that data in the same cluster are generated from a single component density function, so these component density functions are each viewed to describe a cluster. 

The most common model-based approach is a longitudinal latent growth model. These models are based on a longitudinal regression model and assume the outcome variables are from an FMM.  Two examples of latent growth models are the latent class growth analysis (LCGA), which assumes the variance of the random error term varies with time and clusters, and the growth mixture model (GMM), which uses random effects.  An extensive review of latent growth models can be found in \cite{van2020overview,gibbons2010advances,den2020comparison}. However, if random variables in the sequence of outcomes are not continuous and independent, the component density functions of outcomes can be estimated by other statistical models, such as the zero-inflated Poisson model, the censored normal model and the logistic model \citep{Jones01}.
 
An second model-based approach views longitudinal data as high-dimensional cross-sectional data and uses joint distributions for all variables.  \cite{McNicholas10} assume observations at each time point follow a multivariate Gaussian distribution, then apply a modified Cholesky decomposition to the covariance matrix. They show that the component density function is equivalent to the finite mixture model with dynamic model AR(T). \cite{De08} and \cite{Huang18} assume the component joint distributions of a time series are Gaussian with a diagonal covariance, and then use a non-linear model to fit mean vectors. The information from the covariance across time for clustering is ignored. The longitudinal data discussed in these papers are all univariate. 

The third model-based approach is a two-step clustering procedure. In this approach, models are used to estimate trajectories, then the coefficients from these models are used as random variables for clustering. Dynamic regression models, such as auto-regressive (AR), auto-regressive-moving-average (ARMA) and auto-regressive integrated moving average (ARIMA), are often used in the first step of this approach. As an alternative to dynamic regression models, a set of basis functions can be used to fit trajectories, such as B-splines \citep{abraham2003unsupervised}, Fourier basis \citep{serban2005cats}, P-splines \citep{coffey2014clustering}, a Gaussian orthonormal basis \citep{kayano2010functional} and wavelet basis \citep{giacofci2013wavelet}. \cite{wang2016functional} suggest that the selection of a set of basis functions can have an impact on the clustering results.

These three model-based approaches have limitations for multivariate longitudinal data. The first assumes there is a linear relationship between variables and does not consider the autocorrelations of regressors. If the cross-correlations among clusters are similar, this approach will be perform poorly. The second approach considers the autocorrelations between regressors, but the variance matrix is more likely to be singular as the total number of random variables increase. The third approach does not allow the statistical model for estimating trajectories to vary across different features and clusters, which can potentially result in information loss when using a set of coefficients for clustering. The second and third approaches are primarily used for clustering univariate longitudinal data.

The remainder of this paper is organized as follows. Section \ref{sec2} contains relevant background and methodology.  The proposed copula kernel mixture model (CKMM) is introduced for clustering multivariate longitudinal data in Section \ref{sec3}. Section \ref{sec4} contains a simulation study and real data analysis to illustrate the performance of the CKMM in comparison to the DTW and the longitudinal latent growth model. Finally, a discussion of the new clustering method and its performance is included in Section \ref{sec5}.

\section{Background}
\label{sec2}
Let $\mathcal{X}=\{\vecX_n;n=1,\dots,N\}$ denote a balanced longitudinal random sample of size $N$, where a random vector $\vecX=(X_{11},\dots,X_{1T},\dots,X_{D1},\dots,X_{DT})^\top=(\vecX_{1\cdot}^\top,\vecX_{2\cdot}^\top,\dots,\vecX^\top_{D\cdot})^\top$ represents $D$ features being observed $T$ times.  

\subsection{Copulas} 
Each component of a multivariate finite mixture model is a joint density function. In the proposed model, a copula formulation will be used. Sklar's theorem \citep{sklar59} states that any joint cumulative probability function $F(\vecX\mid\mathbf{\Theta})$ can be decomposed into its marginal distributions $F_{1}(x_{11}\mid\boldsymbol{\theta}_{11}),\dots, F_{DT}(x_{DT}\mid\boldsymbol{\theta}_{DT})$ and a $DT$-dimensional copula function $C$ , i.e.,
\begin{equation}
F\left(x_{1},\dots,x_{DT}\mid\boldsymbol{\Theta}\right)=C\left\{F_{11}\left(x_{11}\mid\boldsymbol{\theta}_{11}\right),\dots,F_{DT}\left(x_{DT}\mid\boldsymbol{\theta}_{DT}\right)\mid\boldsymbol{\Theta}_c\right\},
\label{eq:copulacumu}
\end{equation}
for all $x_{11},\dots,x_{DT}$ in the domain of $F$. The parameter $\boldsymbol{\Theta}$ is the collection of marginal parameters $\boldsymbol{\theta}_{11},\dots,
\boldsymbol{\theta}_{DT}$ and copula parameters $\boldsymbol{\Theta}_c$. If the marginal distributions are all continuous, the copula is unique. Taking multiple derivatives of \eqref{eq:copulacumu} with respect to $x_{11},\dots,x_{DT}$ yields the density function $f(x_{11},\dots, x_{DT})$ as follows,
\begin{equation}
     f\left(x_{11},\dots,x_{DT}\right|\boldsymbol{\Theta})=c\left\{F_{11}\left(x_{11}\mid\boldsymbol{\theta}_{11}\right),\dots,F_{DT}\left(x_{DT}\mid\boldsymbol{\theta}_{DT}\right)\mid\boldsymbol{\Theta}_c\right\}\prod_{d=1}^{D}\prod_{t=1}^{T} f_{dt}(x_{dt}\mid\boldsymbol{\theta}_{dt}),
     \label{eq:copuladensity}
\end{equation}
where $c$ is the copula density and $f_{11}\left(x_{11}\mid\boldsymbol{\theta}_{11}\right),\dots,f_{DT}(x_{DT}\mid\boldsymbol{\theta}_{DT})$ are the marginal density functions. 

Many bivariate copula families are not easily generalized to high-dimensions; however, the Gaussian copula is an exception and has a straightforward generalization and estimation in high dimensions. Therefore, the Gaussian copula will be used as the dependence structure for the proposed model. The $DT$-dimensional Gaussian copula density is 
\begin{equation}
    c^{\text{Gauss}}_{\vecR}\left(u_{11},\dots,u_{DT}\right)=\frac{1}{\sqrt{|\vecR|}}\exp{\left\{-\frac{1}{2}\vecq^\top\left(\vecR^{-1}-\mathbf{I}\right)\vecq\right\}},
\label{eq:copulagauss}
\end{equation}
where
\begin{align}
\vecq=\left[ \begin{array}{c}
\Phi^{-1}\left(u_{11}\right),
\dots,
\Phi^{-1}\left(u_{DT}\right)
\end{array} 
\right ]^\top,\nonumber
\end{align}
the matrix $\vecR$ is a copula parameter matrix, $\boldsymbol{I}$ is a $DT\times DT$ identity matrix, $\Phi^{-1}$ is the inverse standard normal cumulative distribution function and is defined for all $\{u_{11},\dots,u_{DT}\}\in(0,1)^{DT}$. 

To estimate the Gaussian copula parameters $\vecR$ and the marginal parameters $\boldsymbol{\theta}_{11},\dots,\boldsymbol{\theta}_{DT}$ in the copula framework, a two-step estimation method is often used rather than maximum likelihood estimation \citep{Joe05,Genest07}. Using \eqref{eq:copuladensity}, the log-likelihood for a sample of size $N$ with observed $DT$-dimensional vectors $\boldsymbol{x_1},\dots,\boldsymbol{x_N}$ is
\begin{equation}
     \ell(\boldsymbol \Theta\!\mid\!\boldsymbol{x})\!=\!\sum_{i=1}^N \log c\left\{F_{11}\left(x_{i11}\mid\boldsymbol{\theta}_{11}\right)\!,\!\dots,F_{DT}\left(x_{iDT}\mid\boldsymbol{\theta}_{DT}\right)\right\}\!+\!\sum_{i=1}^N \sum_{d=1}^{D} \sum_{t=1}^{T}\log f_{dt}(x_{idt}\mid\boldsymbol{\theta}_{dt}).
     \label{eq:copulallik}
\end{equation}
In the first step, the marginal parameters are estimated by maximizing the marginal log-likelihood functions. The second step dictates whether the estimation procedure is semi-parametric or fully parametric. In this step, the original data is transformed to the uniform scale by the probability integral transform using the estimated marginal cumulative distribution functions (CDFs), $\hat{F}_{j}(\cdot)$ for $j=1,\dots,DT$. In the semi-parametric procedure, the empirical CDFs are used for this transformation, whereas the parametric procedure uses the parametric-estimated CDFs, i.e., $F_{j}(\cdot\mid\boldsymbol{\hat\theta}_{j})$. The observations transformed to the uniform scale and inserted into \eqref{eq:copulallik} yields a so-called pseudo-likelihood function, and the dependence parameter $\vecR$ is then estimated by maximizing the first term of this expression.

\subsection{Correlation matrix estimation}
For longitudinal data, the correlation matrix $\vecR$, which parameterizes the Gaussian copula, can be divided into sub-matrices, i.e.,
\begin{align}
    \vecR =\begin{pmatrix}
     \vecR_{11}&\dots&\vecR_{1i}&\dots&\vecR_{1D}\\
     \vdots  & \ddots & \vdots& \ddots & \vdots\\
     \vecR_{i1}&\cdots&\vecR_{ii}&\cdots&\vecR_{iD}\\
     \vdots  & \ddots & \vdots& \ddots & \vdots\\
    \vecR_{D1}&\dots&\vecR_{Di}&\cdots&\vecR_{DD}\\ 
    \end{pmatrix},\nonumber
\end{align}
where the diagonal sub-matrices, $\vecR_{ii}$ for $i=1,\dots,D$, are autocorrelation matrices and the off-diagonals, $\vecR_{ij}$ for $i,j = 1,\dots,D$, $i\neq j$, are cross-correlation matrices. In this model, we assume that the longitudinal data time series are stationary, as non-stationary time series can become stationary after differencing. Due to this assumption, the elements of the correlation matrix $\vecR$ will be defined as a function of the time lags. The structure of $\vecR_{ij}$ is as follows,
\begin{align}
\mathbf{R}_{ij}\!= \!\begin{pmatrix}
r_{ij}(0)&r_{ij}(1)& \cdots & r_{ij}(T-1) \\
r_{ij}(-1) &r_{ij}(0)& \cdots & r_{ij}(T-2)\\
\vdots & \vdots  & \ddots & \vdots\\
r_{ij}(-(T-1))&r_{ij}(-(T-2))&\cdots & r_{ij}(0)
\end{pmatrix},\nonumber
\end{align}
where $r_{ij}(k)$ is the correlations between feature i at time t and feature j at time $(t+k)$ for $k=-(T-1),\dots, (T-1)$  . Therefore, $\mathbf{R}_{ij}$ is a Toeplitz block, i.e., the $D$ diagonal blocks of $\vecR$ are symmetric autocorrelation matrices of the $D$ features, and off-diagonal blocks are asymmetric cross-correlation matrices.  \cite{gray2006toeplitz} gives a circulant matrix to approximate a Toeplitz matrix when the number of dimensions goes to infinity, therefore, the Toeplitz blocks in $\vecR$ will be replaced by circulant blocks, viz. 
\begin{align}
    \hat{\mathbf{R}}_{ij}\!=\!\begin{pmatrix}
r_{ij}(0)&r_{ij}(1)+r_{ij}(-(T-1))& \cdots & r_{ij}(T-1)\!+\!r_{ij}(-1) \\
r_{ij}(-1)\!+\!r_{ij}(T-1) &r_{ij}(0)& \cdots & r_{ij}(T-2)+r_{ij}(2)\\
\vdots & \vdots  & \ddots & \vdots\\
r_{ij}(-(T-1))\!+\!r_{ij}(1)&r_{ij}(-(T-2))\!+\!r_{ij}(2)&\cdots & r_{ij}(0)
\end{pmatrix}.\nonumber
\end{align}
As the eigenvectors of a circulant matrix is a Fourier basis which only relies on the time length and where the time length in this paper is fixed, an eigen-decomposition will be applied to reduce the number of free parameters in $\vecR$. The corresponding eigenvalues and eigenvectors of $\hat{\mathbf{R}}_{ij}$ are:
\begin{align}
    &\lambda_{ij}(m)=\sum_{k=-(T-1)}^{T-1}r_{ij}(k)\exp\left\{2\pi i mk/T\right\},\nonumber\\
    &\boldsymbol{w}_m=\frac{1}{\sqrt{T}}\begin{pmatrix}
    1\\
    \exp\left\{2\pi i m/T\right\}\\
    \vdots\\
    \exp\left\{2\pi i m(T-1)/T\right\}
    \end{pmatrix},\nonumber
\end{align}
respectively. Let $\boldsymbol{W}=\left[\boldsymbol{w}_{0},\dots,\boldsymbol{w}_{T-1}\right] $ and $\boldsymbol{W}^H$ denote the conjugate matrix of $\mathbf{W}$,  then the approximate circulant blocks can be decomposed as $\hat{\mathbf{R}}_{ij}= \mathbf{W}\boldsymbol{\Lambda}_{ij} \mathbf{W}^H$ and $\mathbf{\hat R}$ can be written as:
\begin{align}
{\mathbf{\hat R}}\!=\!\begin{pmatrix}
\mathbf{W} & 0&\cdots&0\\
0 & \mathbf{W}&\cdots&0\\
\vdots &\vdots&\ddots&\vdots\\
0 & 0&\cdots&\mathbf{W}\\
\end{pmatrix}
\!\begin{pmatrix}
\boldsymbol{\Lambda}_{11} & \boldsymbol{\Lambda}_{12}&\cdots&\boldsymbol{\Lambda}_{1D}\\
\boldsymbol{\Lambda}_{21}& \boldsymbol{\Lambda}_{22}&\cdots&\boldsymbol{\Lambda}_{2D}\\
\vdots &\vdots&\ddots&\vdots\\
\boldsymbol{\Lambda}_{D1} & \boldsymbol{\Lambda}_{D2}&\cdots&\boldsymbol{\Lambda}_{DD}\\
\end{pmatrix}
\!\begin{pmatrix}
\mathbf{W}^H & 0&\cdots&0\\
0 & \mathbf{W}^H&\cdots&0\\
\vdots &\vdots&\ddots&\vdots\\
0 & 0&\cdots&\mathbf{W}^H\\
\end{pmatrix},
\label{eq:2}
\end{align}
where the blocks of $\mathbf{\Lambda}$ are diagonal matrices. The number of free parameters in this matrix is reduced from  $TD(TD+1)/2$ to $TD(D+1)/2$, and the approximated matrix, ${\mathbf{\hat R}}$, is closer to the true matrix, ${\mathbf{R}}$, as time length increases. Hence, this approximation is acceptable to reduce computational complexity for large enough $T$.
\subsection{Model Selection}
 The Bayesian information criterion \citep[BIC;][]{schwarz78} is generally used for model selection: 
\begin{equation}
\label{eq:bic}
    \text{BIC}=-2\ell(\hat{\boldsymbol{\Theta}}\mid\boldsymbol{x})+m\log N,\\
\end{equation}
where $\ell(\hat{\boldsymbol{\Theta}}\mid\boldsymbol{x})$ is the maximum log-likelihood value, $m$ is the number of free parameters in the model, and $N$ is the number of observations. The model yielding the lowest BIC score is selected. 

In a finite mixture model context, the BIC can be used to select the number of mixture components. However, in the proposed model, the marginal distributions are estimated by a smoothing method. In this setting, the number of parameters will depend on the sample size, and therefore the BIC assumptions proposed by \cite{blum1977estimation} are violated and the estimator of the number of mixture components is not consistent. Hence,  the number of free parameters in equation~\eqref{eq:bic} is replaced by the number of effective parameters, and this criterion is called adjusted BIC. The number of effective parameters in kernel density estimation is determined by the formula of \cite{mccloud2020determining}. 

An alternative model selection criterion is the normalized entropy criterion (NEC) from \cite{celeux1996entropy}. Note that the expected log-likelihood can be decomposed into two terms:
\begin{align}
    L(G) = \sum_{g=1}^{G}\sum_{n=1}^{N}p_{ng}\log(\pi_{g}f_{g})-\sum_{g=1}^{G}\sum_{n=1}^{N}p_{ng}\log(p_{ng})=C(G) + E(G),\nonumber
\end{align}
where $p_{ng}$ is a posterior probability that subject n belongs to cluster g. Then the normalized entropy is defined as:
\begin{align}
    \text{NEC}(G)= \frac{E(G)}{L(G)-L(1)}, \quad G>1.
\end{align}
For $G>1$, the optimal $G^*$ can be determined by minimizing $\text{NEC}(G)$. To decide whether $G=1$ or $G=G^*$, \cite{celeux1996entropy} re-estimate parameters in the $G^*$ component mixture model but fix mean vectors and mixing proportions in each component at the sample mean and $1/G^*$, respectively, as in the general clustering context, mean vectors in each component are different. The corresponding cross entropy and log-likelihood are denoted by $ \widetilde{E}(1)$ and $ \widetilde{L}(1)$, respectively. If $\text{NEC}(G^*)< \text{NEC}(1)$, the selected number of components is~$G^*$.

However, in this paper, we consider the scenario that the mean values of components from a stationary time series are equal, so the correlation matrix becomes the key set of parameters for clustering. Hence, when we compute the NEC(1), the correlation matrix is fixed to the sample correlation and the other parameters are re-estimated in a $G^*$ component mixture model. As the main contribution of this paper is not to determine the number of clusters, model selections will only be considered in the data analysis.

\section{Methodology}
\label{sec3}
\subsection{Finite Copula Kernel Mixture Model}
In this section, the CKMM is introduced. This model relaxes the distributional assumptions of components in the finite mixture model by using marginal kernel density estimates. The copula kernel density function of $\vecX_d$, $ d=1,\dots,D,$ in cluster $g$ is written as
\begin{align}
    f(\mathbf{x}_n\mid\Theta_{g})=\frac{1}{\sqrt{|\mathbf{R}_{g}|}}\exp\left\{-\frac{1}{2}\mathbf{q}_{n,g}^\top(\mathbf{R}^{-1}_g-\mathbf{I}_{TD})\mathbf{q}_{n,g}\right\}\prod_{d=1}^{D}\prod_{t=0}^{T-1}f_{dt,g}(x_{ndt}),
    \label{eq1}
\end{align}
where $\mathbf{R}_{g}$ is the $g$th copula parameter matrix, which quantifies the dependence of vector $\mathbf{q}_{g}$, $\mathbf{I}_{TD}$ is a $TD\times TD$ identity matrix, and $f_{dt,g}(\cdot)$ is the marginal density function of random variable $X_{dt}$ for $d=1,\dots,D, t=0,\dots,T-1.$ The time series $\{x_{ndt}:t=1,\dots,T\}$ are assumed to be stationary, therefore the marginal densities do not vary with the time index.  
Plugging \eqref{eq:2} to \eqref{eq1} yields
\begin{align}    f(\mathbf{x}_n\mid\boldsymbol\Theta_{g})\!=\!\frac{1}{\sqrt{|\mathbf{\Lambda}_{g}|}}\exp\!\left[\!-\frac{1}{2}\{diag(\boldsymbol{W}^{H})\mathbf{q}_{n,g}\}^H(\mathbf{\Lambda}^{-1}_g\!-\!\mathbf{I}_{TD})\!\{diag(\boldsymbol{W}^H)\mathbf{q}_{n,g}\}\!\right]\prod_{d=1}^{D}\prod_{t=0}^{T-1}f_{d,g}\!(x_{ndt}\!).\nonumber
\end{align}

A permutation matrix $\mathbf{P}$ is multiplied to rearrange the corresponding random variables and make $\boldsymbol{\Lambda}$ become a diagonal block matrix.  The permutation matrix $\mathbf{P}$ is:
\begin{equation}
p_{ij}=\left\{
\begin{aligned}
1,&\quad i=mD+n, \quad j=T(n-1)+(m+1),\\
& \quad m=0,1,\dots,T-1, \quad n=1,2,\dots,D, \\
0,&\quad \text{otherwise,}
\end{aligned}
\right.\nonumber
\end{equation}
and the inverse of matrix $\mathbf{P}$ is $\mathbf{P}^\top$. Then the joint distribution $f(\mathbf{x}_n\mid\boldsymbol\Theta_{g})$ can be written as
\begin{align}
    f(\mathbf{x}_n\mid\Theta_{g})\!=&\!\frac{1}{\sqrt{|\mathbf{\Lambda}_{g}|}}\exp\left[-\frac{1}{2}\{\mathbf{P}diag(\mathbf{W}^H)\mathbf{q}_{n,g}\}^H\{\mathbf{P}(\mathbf{\Lambda}_{g}^{-1}-\mathbf{I}_{TD})\mathbf{P}^\top\}\{\mathbf{P}diag(\mathbf{W}^H)\mathbf{q}_{n,g}\}\right]\nonumber\\
    &\times\prod_{d=1}^{D}\prod_{t=0}^{T-1}f_{d,g}(x_{ndt}),\nonumber
\end{align}
where $(\mathbf{P} \mathbf{\Lambda} \mathbf{P}^\top)^{-1}$ is
\begin{align}
(\mathbf{P} \mathbf{\Lambda} \mathbf{P}^\top)^{-1}=\begin{pmatrix}
\mathbf{C}^{-1}_{0,g} &0& \cdots & 0\\
0 &\mathbf{C}^{-1}_{1,g}& \cdots & 0\\
\vdots & \vdots  & \ddots & \vdots\\
0&0&\cdots & \mathbf{C}^{-1}_{T-1,g}
\end{pmatrix},\nonumber
\quad \mathbf{C}_{j,g}=\begin{pmatrix}
\lambda_{11,g}(j) & \cdots & \lambda_{1D,g}(j)\\
\lambda_{21,g}(j) & \cdots & \lambda_{2D,g}(j)\\
\vdots   & \ddots & \vdots\\
\lambda_{D1,g}(j)&\cdots & \lambda_{DD,g}(j)
\end{pmatrix}.
\end{align}
In this form, the joint density function can be decomposed to the product of marginal densities: 
\begin{align}
     f(\mathbf{x}_n\mid\Theta_{g})=&\prod_{j=0}^{T-1}\left\{\frac{1}{\sqrt{|\mathbf{C}_{j,g}|}}\exp\left\{-\frac{1}{2}(diag(\mathbf{w}_{j}^H)_{D\times D}\mathbf{q}_{n,g})^H(\mathbf{C}^{-1}_{j,g}-\mathbf{I}_{D})(diag(\mathbf{w}_{j}^H)_{D\times D}\mathbf{q}_{n,g})\right\}\right.\nonumber\\
    &\left.\times\prod_{d=1}^{D}f_{d,g}(x_{ndj})\right\},
\label{eq:3}
\end{align}
where
\begin{align}
    \mathbf{w}^H_{j} = \frac{1}{\sqrt{T}}\left(\exp\left\{-\frac{2\pi j\times 0i}{T}\right\},\exp\left\{-\frac{2\pi j\times 1i}{T}\right\},\dots,\exp\left\{-\frac{2\pi j\times (T-1)i}{T}\right\}\right).\nonumber
\end{align}

As the dependence parameters of the copula function are estimated using the marginal cumulative probabilities, an incorrect parametric assumption on marginal distributions may lead to estimation bias and reduced performance of the CKMM in a fully parametric estimation approach. Hence, marginal density functions will be estimated via a kernel smoothing method. The finite copula kernel mixture model can be defined via the component density functions.
\begin{align}
    f(\mathbf{x}_{n}\mid\boldsymbol\Theta)=\sum_{g=1}^{G}\pi_gf(\mathbf{x}_{x}\mid Z_{n}=g,\boldsymbol\Theta_g),
    \label{eq:4}
\end{align}
where $\mathbf{\Theta}=\left\{\pi_{g},\mathbf{C}_{j,g},f_{d,g},\vecq_{g}\right\}_{1\leq g\leq G, 1\leq d \leq D, 0\leq j \leq T-1}$ is the parameter space and $Z_{n}$ is a latent variable representing the mixture component for $\mathbf{x}_n$. The approach is a semi-parametric, where the marginal distributions do not change over time due to a stationary assumption.  

\subsection{Estimation}

As the variable $Z_n$ is not observable and the parameter space includes both functional and Euclidean parameters, a generalized EM (GEM) algorithm will be used to estimate the CKMM. The GEM algorithm is a generalization of the EM algorithm where each M-step only requires a non-decrease of the objective function rather than a maximization. The GEM can handle the difficult problem where the estimation of the Euclidean parameters relies on the functional estimator of marginal density functions.  As the GEM algorithm may converge to local maxima, different initializations are considered to ensure the global maximum is reached. 

To estimate the marginal density functions, a nonlinear, concave smoothing operator $\mathcal{N}$ \citep{levine2011maximum} is used for the density functions, which is shown as below:
\begin{align}
    \mathcal{N} f_{d,g} (x_{ndt}) = \exp \int_{-\infty}^{\infty} K_{h_{d,g}}(x_{ndj},u)\log(f_{d,g}(u))du.\nonumber
\end{align}

In this setting, $\left\{\mathbf{x}_n,z_n\right\}$ is considered the complete data and the resulting expected complete-data log-likelihood is :
\begin{align}
\label{eq:Q} &E\left(\!\ell_{\boldsymbol{X}}\!\right)\!=\sum_{n=1}^{N}\sum_{g=1}^{G}p_{ng}\left\{\sum_{j=0}^{T-1}\left[-\frac{1}{2}(diag(\mathbf{w}_{j}^H)_{D}\mathbf{q}_{n,g})^H(\mathbf{C}^{-1}_{j,g}-\mathbf{I}_{D})(diag(\mathbf{w}_{j}^H)_{D}\mathbf{q}_{n,g})\right.\right.\nonumber\\
    &\left.\left.-\frac{1}{2}\log(|\mathbf{C}_{j,g}|)\right]\!+\sum_{j=0}^{T-1}\sum_{d=1}^D\int_{-\infty}^{\infty} K_{h_{d,g}}(x_{ndj},u)\log(f_{d,g}(u))du+\log \pi_g-\log p_{ng}\right\},
\end{align}
 where $p_{ng}$ is a posterior probability $P(Z_n=g\mid \mathbf{X}_n=\mathbf{x}_n)$, the kernel function $K$ is the Gaussian kernel function, and $h_{d,g}$ is bandwidth for feature $d$ and cluster $g$. The cumulative values $\mathbf{q}$ are computed based on the estimated marginal density function. 
 The E-step is similar to a standard EM algorithm, the M-step at $(k+1)$th iteration includes the following four sub steps. Below is an outline of the steps of the GEM algorithm.
 
\paragraph{E-step} Compute the posterior probabilities $\hat{p}_{ng}$ at $(k+1)$th iteration.
\begin{align}
\label{eq:post}
    \hat{p}^{(k+1)}_{ng}=\frac{\hat{\pi}^{(k+1)}_g\mathbf{c}(\hat{\vecq}^{(k+1)}_{ng},\hat{\vecC}^{(k+1)}_{g})\prod_{d=1}^{D}\prod_{j=0}^{T-1}\exp{\int_{-\infty}^{\infty}} K_{h^{(k+1)}_{d,g}}(x_{ndj},u)\log \hat{f}^{(k+1)}_{d,g}(u)du}{\sum^{G}_{g=1} \hat{\pi}^{(k+1)}_g\mathbf{c}(\hat{\vecq}^{(k+1)}_{ng},\hat{\vecC}^{(k+1)}_{g})\prod_{d=1}^{D}\prod_{j=0}^{T-1}\exp{\int_{-\infty}^{\infty}} K_{h^{(k+1)}_{d,g}}(x_{ndj},u)\log \hat{f}^{(k+1)}_{d,g}(u)du}.
\end{align}
 
\paragraph{M-step 1} Estimate functions $f^{(k+1)}_{d,g}$
 \begin{align}
     &\hat{f}^{(k+1)}_{d,g}(u)= \arg\max_{f_{d,g}}\left(\int_{-\infty}^{\infty}\sum_{n=1}^{N}\sum_{j=0}^{T-1}p^{(k+1)}_{ng}K_{h_{d,g}}(x_{ndj},u)\log f_{d,g}(u)du\right),\nonumber \end{align}
such that $$\int_{-\infty}^{\infty}\hat{f}^{(k+1)}_{d,g}(u)du=1.$$
 Thus, the kernel density estimate is 
\begin{align}
\label{kde}
    \hat{f}^{(k+1)}_{d,g}(u)=\frac{1}{T\sum_{n=1}^{N}p^{(k+1)}_{ng}}\sum_{n=1}^{N} \sum_{j=0}^{T-1} p^{(k+1)}_{ng}K_{h_{d,g}}(x_{ndj},u).
\end{align}

\paragraph{M-step 2} Choose bandwidth $h_{d,g}$ and calculate $\vecq_{g}$. An analytical solution for bandwidth cannot be obtained by maximizing the expected log-likelihood, as both $\vecq_{ng}$ and the marginal density estimates are affected by the bandwidth.  Also, the pseudo log-likelihood with respect to bandwidth is not necessarily concave, so numerical methods, such as gradient descent, are invalid for finding the optimum value.  Hence, in this step, instead of computing the conditional maximum values of bandwidth, the bandwidths from a given interval which increase the pseudo expected complete-data log-likelihood function will be chosen. Searching starts at the bandwidths from the last iteration, and then a small $\Delta h$ is added to $h^{(k)}_{d,g}$, denoted by $h^{(k+1,1)}_{d,g}$. The bandwidth sequence at $(k+1)$th iteration will be
\begin{align}
\label{eq:band}
    \hat{h}^{(k+1,i+1)}_{d,g} \!=\!  \hat{h}^{(k+1,i)}_{d,g}\! \!+\!\! \eta\frac{E\!\left(\!\ell_{\boldsymbol{X}}(h^{(k+1,i)}_{d,g},\vecq^{(k+1,i)}_{n,g})\mid\hat{f}^{(k+1)}\!\right)\!\!-\!\!E\!\left(\!\ell_{\boldsymbol{X}}(h^{(k+1,i-1)}_{d,g},\vecq^{(k+1,i-1)}_{n,g})\mid\hat{f}^{(k+1)}\!\right)}{h^{(k+1,i)}_{d,g}\!-\!h^{(k+1,i-1)}_{d,g}}.
\end{align}
where $\eta$ is a learning rate, and $i=1,2,\dots$. If $\hat{h}^{(k+1,i)}_{d,g}$ exceeds the range of bandwidth or the change of pseudo log-likelihood is smaller than some threshold, the iterations will stop.  

\paragraph{M-step 3} Estimate the correlation matrices by maximizing conditional pseudo log-likelihood subject to constraint on $\hat{h}^{(k+1)},\hat{\vecq}^{(k+1)},\hat{f}^{(k+1)},\hat{p}^{(k)}_{ng}$. The correlation estimators are:
\begin{align}
\label{eq:c}\hat{\vecC}^{(k+1)}_{ j,g}=\frac{1}{\sum_{n=1}^{N}\hat{p}_{ng}^{(k)}}\sum_{n=1}^{N}\hat{p}_{ng}^{(k)}\left(diag(\mathbf{w}_{j}^H)_D\hat{\vecq}^{(k+1)}_{n,g}\right)\left(diag(\mathbf{w}_{j}^H)_D\hat{\vecq}^{(k+1)}_{n,g}\right)^{H}
\end{align}
{for $j=0,...,T-1$ and $g=1,\dots,G$.}

\paragraph{M-step 4} Estimate the prior probabilities $\boldsymbol\pi$, given $\hat{\mathbf{C}}^{(k+1)},\hat{h}^{(k+1)},\hat{\vecq}^{(k+1)},\hat{f}^{(k+1)},\hat{p}^{(k)}_{ng}$ viz.
\begin{align}
    \label{eq:prior}\hat{\pi}^{(k+1)}_g=\frac{\sum_{n=1}^{N}\hat{p}^{(k)}_{ng}}{N}
\end{align}
The CKMM clustering method is summarized in Algorithm~\ref{CKMM}, and the threshold $\epsilon$ is set to $1\times 10^{-5}$ for the analyses in Section~\ref{sec4}. 

\begin{algorithm}[ht]
	\renewcommand{\algorithmicrequire}{\textbf{Input:}}
	\renewcommand{\algorithmicensure}{\textbf{Output:}}
\caption{Estimating the copula kernel mixture model.}\label{CKMM}
	\begin{algorithmic}[1]
	\REQUIRE sample data $\boldsymbol{X}=\left\{\boldsymbol{x}_{1},\dots,\boldsymbol{x}_{n}\right\}$ and the number of clusters $G$
	\ENSURE clustering labels, estimates $\hat{\vecC},\hat{f},\hat{\pi},\hat{\vecq},\hat{h}$ and pseudo  log-likelihood value.
	\STATE Initialize labels and parameters by $K$-means
	\REPEAT
	\STATE $k \leftarrow k + 1$
	\STATE \textbf{E-step}:
		\FOR{$\boldsymbol{x}_{n} \in \boldsymbol{X}$} 
		\FOR{$g \gets 1$ to $G$}
            \STATE compute \textit{a posteriori} probability $\hat{p}_{n,g}^{(k+1)}$
        \ENDFOR
        \STATE compute pseudo log-likelihood value $E(\ell(\mathbf{\Theta}^{(k+1)}))$ 
        \STATE determine the clustering label based on the \textit{a posteriori} probability
		\ENDFOR
		\STATE \textbf{M-step}:
		\FOR{$g \gets 1$ to $G$} 
		\FOR{$d \gets 1$ to $D$}
        \STATE \textbf{M-step1}: update marginal density function estimates $\hat{f}_{d,g}^{(k+1)}$ 
        \STATE \textbf{M-step2}: update bandwidths $\hat{h}_{d,g}^{(k+1)}$ and $\hat{\vecq}_{g}^{(k+1)}$ 
        \ENDFOR
         \STATE \textbf{M-step3}: update $\hat{\vecC}_{g}^{(k+1)}$ 
        \STATE \textbf{M-step4}: update prior probabilities $\pi_{g}^{(k+1)}$ 
	
		\ENDFOR
		\UNTIL {$|\ell(\Theta^{(k+1)})-\ell(\boldsymbol{\Theta}^{(k)})|/|\ell(\boldsymbol{\Theta}^{(k+1)})| < \varepsilon $}
	\end{algorithmic}
\end{algorithm}

\section{Simulation Study}
\label{sec4}
In this section, the clustering performance of the proposed CKMM will be compared with two common longitudinal clustering methods: $K$-means clustering using DTW distance and the LCGA.  

 Six scenarios of bivariate longitudinal data will be considered with time lengths varying from 20 to 50 in each scenario. The number of clusters, sample size, and prior probabilities are the same for each scenario: 2, 100 and 0.4, respectively. To avoid non-representative performance results, 100 samples are generated for each scenario. The six scenarios are summarized as below$\colon$
\begin{itemize}
    \item [S1]: The correlation of the two features does not vary across clusters and is fixed at 0.
    \item [S2]: The correlation of the two features  does not vary across clusters and is fixed at 0.25.
    \item [S3]: The correlation of the two features  does not vary across clusters and is fixed at 0.5.
    \item [S4]: The correlation of the two features varies across clusters, and the correlations are 0 and 0.25, respectively.
    \item [S5]: The correlation of the two features varies across clusters, and the correlations are 0 and 0.5, respectively.
    \item [S6]: The correlation of the two features varies across clusters, and the correlations are 0.25 and 0.5, respectively.
\end{itemize}

\subsection{Data Generation}
Suppose that the time series of each feature is generated by a moving average (MA) model. As the autocorrelation function cuts off after lag $q$ in MA($q$), most elements in the autocovariance matrix are 0 and an unbiased estimator of the variance matrix will be  a singular matrix when time length is long. The non-invertibility of the variance will influence the clustering performance of the Gaussian mixture model. The correlation between two features will be simulated by a vector autoregressive (VAR) model. The matrix form of a bivariate VAR model can be written as: 
\begin{align}
    &\text{VAR}_{\text{cluster} 1}:    \begin{pmatrix}
    x_{1T}\\
    x_{2T}
    \end{pmatrix}=\begin{pmatrix}
   \theta_{11}&0\\
    0&\theta_{21}\\
    \end{pmatrix}\begin{pmatrix}
    \epsilon_{1T-1}\\
    \epsilon_{2T-1}
    \end{pmatrix}+\begin{pmatrix}
    \epsilon_{1T}\\
    \epsilon_{2T}
    \end{pmatrix},\nonumber\\
     &\text{VAR}_{\text{cluster} 2}:    \begin{pmatrix}
    x_{1T}\\
    x_{2T}
    \end{pmatrix}=\begin{pmatrix}
    \epsilon_{1T}\\
    \epsilon_{2T}
    \end{pmatrix}+\begin{pmatrix}
    \theta_{11}&0\\
    0&\theta_{21}
    \end{pmatrix}\begin{pmatrix}
    \epsilon_{1T-1}\\
    \epsilon_{2T-1}
    \end{pmatrix}+\begin{pmatrix}
  \theta_{12}&0\\
    0&\theta_{22}
    \end{pmatrix}\begin{pmatrix}
    \epsilon_{1T-2}\\
    \epsilon_{2T-2}
    \end{pmatrix},\nonumber
\end{align}
where $x_{iT}$ is the observed value of $i$th feature at time point $T$, and the random error $\epsilon_{iT}$ is independent of the time indices. The autocovariance matrices for features will be computed based on the following rules.
\begin{align}
  &\text{Cluster} 1: \text{cov}(X_{iT},X_{iT})=(1+\theta^2_{i1})\text{var}(\epsilon_{iT}),\nonumber\\
  &\quad\qquad  \qquad\quad\ \text{cov}(X_{iT},X_{iT-1})=\theta_{i1}\text{var}(\epsilon_{iT}),\nonumber\\
  &\quad\qquad  \qquad\quad\ \text{cov}(X_{iT},X_{iT-t})=0 \qquad \qquad \qquad \text{t}\geq 2,\quad \text{i} = 1,2.\nonumber\\
    &\text{Cluster} 2: \text{cov}(X_{iT},X_{iT})=(1+\theta^2_{i1}+\theta^2_{i2})\text{var}(\epsilon_{iT}),\nonumber\\
  &\quad\qquad  \qquad\quad\ \text{cov}(X_{iT},X_{iT-1})=(\theta_{i1}+\theta_{i1}\theta_{i2})\text{var}(\epsilon_{iT}),\nonumber\\
  &\quad\qquad  \qquad\quad\ \text{cov}(X_{iT},X_{iT-2})=\theta_{i2}\text{var}(\epsilon_{iT}),\nonumber\\
  &\quad\qquad  \qquad\quad\ \text{cov}(X_{iT},X_{iT-t})=0 \qquad \qquad \qquad \text{t}\geq 3,\quad \text{i} = 1,2.\nonumber
 \end{align}
 The cross-correlation matrices of two features depend on the correlations of the random error terms. Let $A_1$, $A_2$ and $B_2$ represent coefficient matrices in $\text{VAR}_{\text{cluster} 1}$ and $\text{VAR}_{\text{cluster} 2}$. The covariance matrices of $X_{1T}$ and $X_{2T}$ in the clusters are: 
 \begin{align}
  &\text{Cluster 1}: \text{cov}(X_{1T},X_{2T})=\text{E}(\Vec{\epsilon}_{T}\Vec{\epsilon}^\top_{T})+\text{E}(A_1\Vec{\epsilon}_{T-1}\Vec{\epsilon}^\top_{T-1}A^\top_{1}),\nonumber\\
  &\text{Cluster 2}: \text{cov}(X_{1T},X_{2T})=\text{E}(\Vec{\epsilon}_{T}\Vec{\epsilon}^\top_{T})+\text{E}(A_1\Vec{\epsilon}_{T-1}\Vec{\epsilon}^\top_{T-1}A^\top_{1})+\text{E}(B_1\Vec{\epsilon}_{T-1}\Vec{\epsilon}^\top_{T-1}B^\top_{1}).\nonumber
 \end{align}
 The correlations of $X_{1T}$ and $X_{2T}$ in the clusters are: 
 \begin{align}
       &\text{Cluster 1}: \text{cor}(X_{1T},X_{2T})= \frac{(1+\theta_{11}\theta_{21})\text{cov}(\epsilon_{1T},\epsilon_{2T})}{\sqrt{(1+\theta^2_{11})\text{var}(\epsilon_{1T})}\sqrt{(1+\theta^2_{21})\text{var}(\epsilon_{2T})}},\nonumber\\
       &\text{Cluster 2}: \text{cor}(X_{1T},X_{2T})= \frac{(1+\theta_{11}\theta_{21}+\theta_{12}\theta_{22})\text{cov}(\epsilon_{1T},\epsilon_{2T})}{\sqrt{(1+\theta^2_{11}+\theta^2_{12})\text{var}(\epsilon_{1T})}\sqrt{(1+\theta^2_{21}+\theta^2_{22})\text{var}(\epsilon_{2T})}}.\nonumber
 \end{align}
 When correlations of features are known, the correlations of the random errors can be computed based on the above equations. The cross-variance matrices are determined by the following equations:
 \begin{align}
     &\text{Cluster 1}: \text{cov}(X_{1T},X_{2T-1})=\theta_{11}\text{cov}(\epsilon_{1T},\epsilon_{2T}),\nonumber\\
     &\qquad \qquad \quad \text{cov}(X_{1T-1},X_{2T})=\theta_{21}\text{cov}(\epsilon_{1T},\epsilon_{2T}),\nonumber\\
     &\qquad \qquad \quad \text{cov}(X_{1T-t},X_{2T})=\text{cov}(X_{1T},X_{2T-t})=0 \qquad t \geq 2, \nonumber\\
      &\text{Cluster 2}: \text{cov}(X_{1T},X_{2T-1})=(\theta_{11}+\theta_{12}\theta_{21})\text{cov}(\epsilon_{1T},\epsilon_{2T}),\nonumber\\
     &\qquad \qquad \quad \text{cov}(X_{1T-1},X_{2T})=(\theta_{21}+\theta_{11}\theta_{22})\text{cov}(\epsilon_{1T},\epsilon_{2T}),\nonumber\\
     &\qquad \qquad \quad \text{cov}(X_{1T},X_{2T-2})= \theta_{12}\text{cov}(\epsilon_{1T},\epsilon_{2T}),\nonumber\\
     &\qquad \qquad \quad \text{cov}(X_{1T-2},X_{2T})= \theta_{22}\text{cov}(\epsilon_{1T},\epsilon_{2T}),\nonumber\\
     &\qquad \qquad \quad \text{cov}(X_{1T-t},X_{2T})=\text{cov}(X_{1T},X_{2T-t})=0 \qquad t \geq 3. \nonumber
 \end{align}
The specific parameter values used for covariance matrix generation are summarized in Table~\ref{tab:datag}. Note that feature 1 is dissimilar between clusters as it is generated by MA(1) and MA(2), respectively, from each cluster and coefficients of first lag have opposite signs. In contrast, feature 2 is more similar in between clusters, as it is generated from MA(1) models with similar coefficients. The correlation matrices $\textbf{R}_{1}$ and $\textbf{R}_{2}$ can be calculated using the parameters in Table~\ref{tab:datag}. 
\begin{table}[H]
\setlength\tabcolsep{6pt}
\centering
\caption{Covariance matrix generation, where $\rho$ and $\theta$ represent correlations and coefficients, respectively.}
\begin{tabular}{c|cccccccc}
\hline
Scenarios           & Cluster  & $\theta_{11}$ & $\theta_{12}$ & $\theta_{21}$ & $\theta_{22}$ & $\rho _{cross}$ & $\rho _{\epsilon}$ & $\text{var}_{\epsilon}$ \\ \hline
\multirow{2}{*}{S1} & cluster1 & -0.2679     & 0           & 0.6268      & 0           & 0           & 0                  & 1                       \\
                    & cluster2 & 0.2532      & 0.0533      & 0.5000      & 0           & 0           & 0                  & 1                       \\ \hline
\multirow{2}{*}{S2} & cluster1 & -0.2679     & 0           & 0.6268      & 0           & 0.25        & 0.3671             & 1                       \\
                    & cluster2 & 0.2532      & 0.0533      & 0.5000      & 0           & 0.25        & 0.2562             & 1                       \\ \hline
\multirow{2}{*}{S3} & cluster1 & -0.2679     & 0           & 0.6268      & 0           & 0.5         & 0.7342             & 1                       \\
                    & cluster2 & 0.2532      & 0.0533      & 0.5000      & 0           & 0.5         & 0.5125             & 1                       \\ \hline
\multirow{2}{*}{S4} & cluster1 & -0.2679     & 0           & 0.6268      & 0           & 0           & 0                  & 1                       \\
                    & cluster2 & 0.2532      & 0.0533      & 0.5000      & 0           & 0.25        & 0.2562             & 1                       \\ \hline
\multirow{2}{*}{S5} & cluster1 & -0.2679     & 0           & 0.6268      & 0           & 0           & 0                  & 1                       \\
                    & cluster2 & 0.2532      & 0.0533      & 0.5000      & 0           & 0.5         & 0.5125             & 1                       \\ \hline
\multirow{2}{*}{S6} & cluster1 & -0.2679     & 0           & 0.6268      & 0           & 0.25           & 0.3671                  & 1                       \\
                    & cluster2 & 0.2532      & 0.0533      & 0.5000      & 0           & 0.5         & 0.5125             & 1                       \\ \hline                    
\end{tabular}
\label{tab:datag}
\end{table}

\begin{table}[H]
\setlength\tabcolsep{6pt}
\centering
\caption{Marginal distributions.}
\begin{tabular}{c|cc}
\hline
         & feature1    & feature2          \\ \hline
cluster1 & Normal(1,1.0718) & t-distribution(df=7.0908) \\
cluster2 & Normal(1,1.0669) & t-distribution(df=10)     \\ \hline
\end{tabular}
\label{tab:marginalg}
\end{table}

To illustrate that CKMM could also fit data from a non-Gaussian distribution, the marginal distribution of the second feature is a t-distribution. The standard deviations of the normal distribution and the degrees of freedom of the t-distribution are computed based on diagonal elements in the respective covariance matrix. The data generation process is shown as below.
\begin{enumerate}
    \item Sample 100 labels from Bernoulli distribution with $p = 0.4$.
    \item Sample quantile vectors from multivariate Gaussian distribution with mean 0 and variance $R_1$ or $R_2$, where the variance is determined by the labels.
    \item Convert the quantile vectors to the time series vectors based on the marginal distributions as shown in Table~\ref{tab:marginalg}.
\end{enumerate}
\subsection{Clustering Analysis}
To avoid convergence to a local optimum, ten initializations will be generated from $K$-means. Initial bandwidths, correlation
matrices, kernel density estimates, and \textit{a priori} probabilities are computed based on these initial labels. In this study, we assume the number of clusters is known ($G=2$). The results with maximum observed log-likelihood will be selected as final results. The adjusted Rand index (ARI) of \cite{hubert85} will be used to evaluate clustering performance. An ARI value of 1 corresponds to perfect class agreement whereas 0 is the expected ARI under random classification. 
 \begin{figure}[!htb]
 \centering
 \subfigure[]{
  \centering
  \includegraphics[width=0.22\linewidth,height=3.5cm]{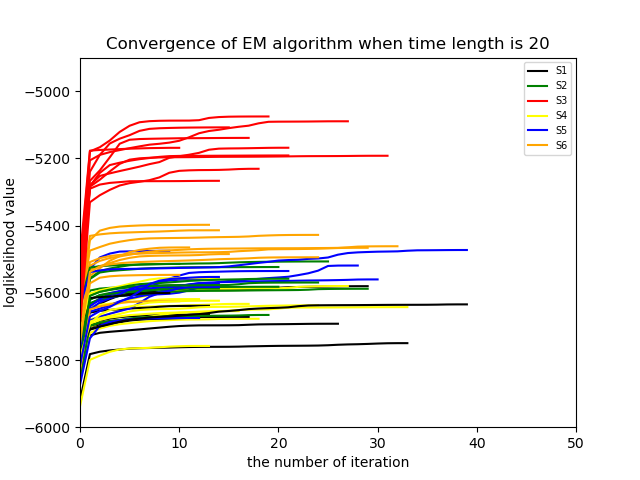}
  \label{subfig:T20}}
 \subfigure[]{
  \centering
  \includegraphics[width=0.22\linewidth,height=3.5cm]{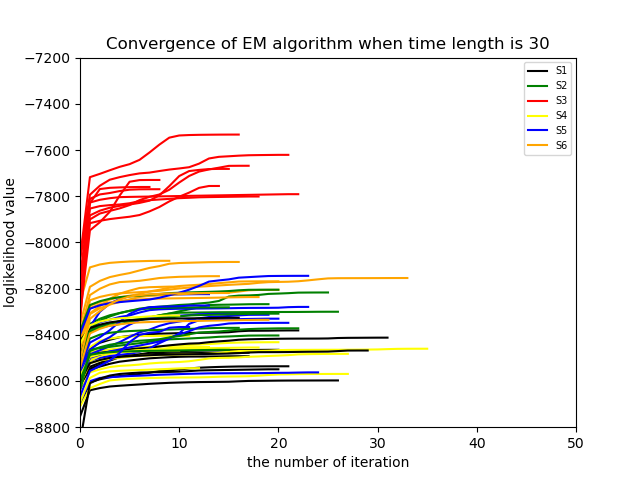}
  \label{subfig:T30}}
 \subfigure[]{
  \centering
  \includegraphics[width=0.22\linewidth,height=3.5cm]{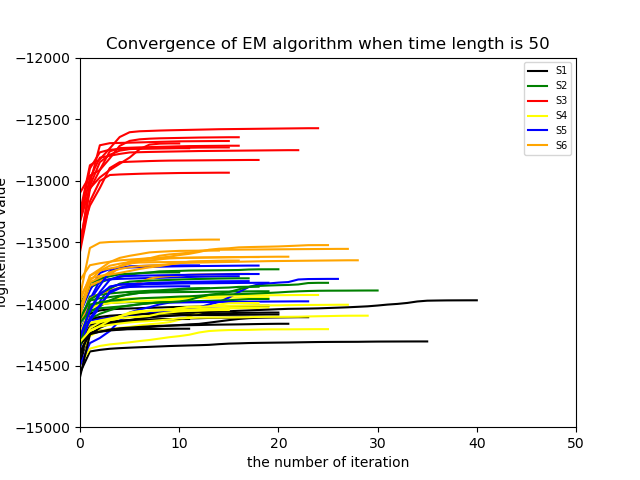}
  \label{subfig:T50}}
 \subfigure[]{
  \centering
  \includegraphics[width=0.22\linewidth,height=3.5cm]{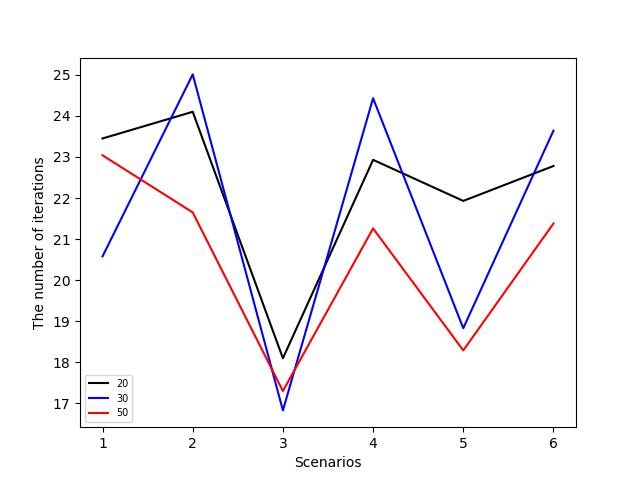}
  \label{subfig:meanit}}
\caption{(a)--(c) The log-likelihood values computed at each iteration of the GEM algorithm for first ten datasets of each simulation scenario (colored) with time lengths 20, 30, and 50, respectively. (d) The average number of iterations on 100 datasets for each simulation scenario with time lengths 20, 30, and 50 (colored).}
\label{fig:loglike}
\end{figure}

The log-likelihood values of the first ten data sets for each scenario and time length are used as examples to show the convergence of EM algorithm in CKMM, shown in Figure~\ref{subfig:T20}--\ref{subfig:T50}. The log-likelihood values of S3 are the highest among the three different time lengths, as the determinants of correlation matrices in S3 are smaller than those from the other scenarios. The log-likelihood values decrease as time lengths increase, because data in high-dimensional space tend to be sparse and corresponding density values will be small. It is clear that  log-likelihood values of each scenarios and time lengths converge to fixed values after several iterations, and the log-likelihood values increase monotonically with the number of iterations. Thus, the GEM algorithm for the CKMM converges and can therefore be used to estimate the parameters of CKMM.  The average number of iterations for each scenario and time length is plotted in Figure~\ref{subfig:meanit} to display the convergence speed of EM algorithm. 

\begin{figure}[!htb]
\centering
\includegraphics[width=0.75\linewidth,height=5.5cm]{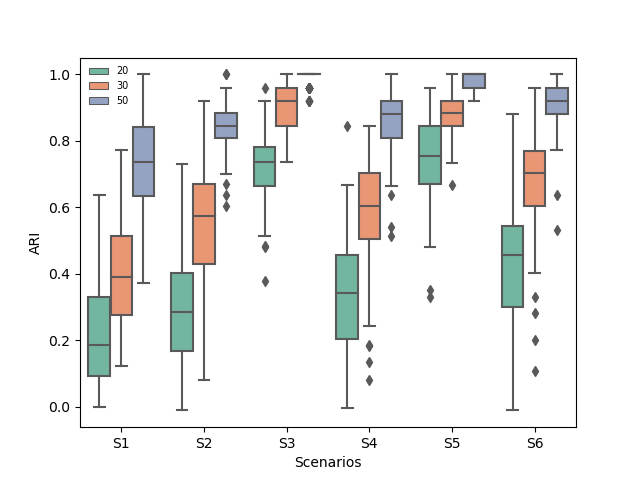}
\caption{Boxplots of ARI results from the CKMM across the six simulation scenarios, where the colors represent three time lengths. One hundred datasets were generated from each scenario and for each time length. Some outliers exist because GEM may converge to a local optimum. }
\label{fig:ARI values}
\end{figure}

Figure~\ref{fig:ARI values} and Table~\ref{Tab:ari} compare the ARI values on 100 simulated datasets for each scenario and time length. In terms of time length, increasing time lengths in a fixed scenario yields an average increase in clustering performance. For example,  the mean ARI of time length 50 for S1 is 0.7218 compared to 0.2186 for a time length of 20. The first three groups of boxplots in Figure~\ref{fig:ARI values} show that ARI increases, on average, as the correlation of features increases even if the correlations is the same in both clusters. The mean ARI for a time length of 50 increases from 0.7218 for S1 to 0.9884 for S3. The last three groups in comparison with the first three groups illustrate that if the correlations vary with clusters, the performance improves, especially for time lengths of 20 and 30. Also, the difference of correlations in two cluster influences the performance. The ARI of S5 is the highest among the last three scenarios, and the difference of correlations in two clusters is also largest (0.5).

\begin{table}[!htb]
\setlength\tabcolsep{3pt}
\centering
\caption{ARI means with corresponding standard errors in parentheses over multiple initializations.}
\begin{tabular}{c|c|c|c|c|c}
\hline
Scenarios           & Time length              &CKMM   &DTW &LCGA  &Baseline                \\ \hline
\multirow{3}{*}{S1} & 20   &\textbf{0.2186}(0.1521)&  0.0603(0.0407)  &0.0179 (0.0387)
& 0.5799(0.1039) \\
                    & 30      &\textbf{0.4033}(0.1595) &0.0179 (0.0387)  & 0.0133 (0.0324)               
&0.7312(0.0987) \\
                    & 50     &\textbf{0.7218}(0.1323)& 0.1189(0.0741)&0.0175 (0.0337)
&0.8880(0.0707)\\ \hline
\multirow{3}{*}{S2} & 20 &\textbf{0.2849}(0.1557)&0.0530(0.0330)&0.0173 (0.0303)
&0.6535(0.0947) \\
                    & 30          &\textbf{0.5516}(0.1662)                    &0.0661(0.0477)&0.0166 (0.0323)
&0.7852(0.0816) \\
                    & 50          & \textbf{0.8475}(0.0737)&0.1213(0.0728)&0.0181 (0.0362)
&0.9334(0.0530)\\ \hline
\multirow{3}{*}{S3} & 20         &\textbf{0.7229}(0.1000)&0.0443(0.0361)&0.0145 (0.0323)
&0.8845(0.0610)\\
                    & 30       &\textbf{0.8982}(0.0577)&
                    
0.0620(0.0470)& 0.0133 (0.0290)&0.9583(0.0413) \\
                    & 50           &\textbf{0.9884}(0.0220)&
0.1074(0.0672)&0.0136 (0.0261) &0.9952(0.0130) \\ \hline
\multirow{3}{*}{S4} & 20          &\textbf{0.3389}(0.1642)& 0.0534(0.0390)&0.1353 (0.1042)
&0.6752(0.0918)\\
                    & 30         &\textbf{0.5838}(0.1621) &0.0677(0.0486)&0.1855 (0.1106)
&0.7885(0.0812)\\
                    & 50        & \textbf{0.8511}(0.0840)&0.1225(0.0743)&0.3379 (0.1365)
&0.9314(0.0526)\\ \hline
\multirow{3}{*}{S5} & 20        &\textbf{0.7468}(0.1209)&0.0540(0.0379)&0.5828 (0.1050)
&0.8653(0.0708) \\
                    & 30        &\textbf{0.8907}(0.0619)&
0.0609(0.0526)& 0.7065 (0.1013)&0.9320(0.0460) \\
                    & 50          & \textbf{0.9847}(0.0238)&0.1061(0.0625)&0.8927 (0.0651) 
&0.9908(0.0202)\\ \hline
\multirow{3}{*}{S6} & 20         &\textbf{0.4198}(0.1793)&0.0517(0.0415)&0.0079 (0.0117)
&0.7119(0.0861) \\
                    & 30   &\textbf{0.6666}(0.1488)&
0.0632(0.0453)&0.0010 (0.0042)&0.8384(0.0641) \\
                    & 50        & \textbf{0.9146}(0.0721)&0.1171(0.0785)&0.4418 (0.1299)
&0.9591(0.0395)\\ \hline
\end{tabular}
\label{Tab:ari}
\end{table}

 $K$-means clustering using DTW distance is performed by the function \texttt{TimeSeriesKMeans} from \texttt{tslearn} in {\sf Python}, with the parameter metric specified as \texttt{dtw}. To fit the LCGA, the first feature is used for the dependent variable and the second feature is the covariate. The function \texttt{stepFlexmix} from the \texttt{flexmix} {\sf R} package \citep{R20} is used to fit the LCGA. The ARI results are shown in Table~\ref{Tab:ari}.

The average ARI of $K$-means based on DTW distance and CKMM both show that ARI increases on average with time length. Although the DTW distance is designed for time series, distance-based clustering has the poorest overall performance among these three methods. Thus, $K$-means based on DTW distance does not appear to be suitable for stationary multivariate time series with the same mean values. The LCGA clusters longitudinal data via cross-correlations. If the correlations in two cluster are the same, the model performance will be poor, such as the first three scenarios. If the correlations are different, the performance increases significantly. However, the mean ARI of LCGA is still smaller than that of CKMM. Thus, the CKMM has shown better average performance than the alternative methods across all simulation scenarios considered.

The baseline column in Table~\ref{Tab:ari} is the mean ARI when the parameters of the CKMM set equal to their true values. In the table we find that the changes in the ARIs of the CKMM are consistent with the changes of the baseline, but the ARIs of the CKMM are always smaller than baseline values. The differences between the CKMM and the baseline become smaller as time length and correlation increases. However, the standard deviation of the ARIs for the CKMM is largest among the three methods.
 
\subsection{Estimator Analysis}
In the CKMM, the parameters consist of the \textit{a priori} probabilities, correlation matrices, and bandwidths. Based on the estimated bandwidths, we can obtain kernel density functions to estimate the marginal distributions. 
\begin{figure}[!htp]
\centering
 \subfigure[KDE of F1C1]{
  \centering
  \includegraphics[width=0.22\linewidth,height=3.5cm]{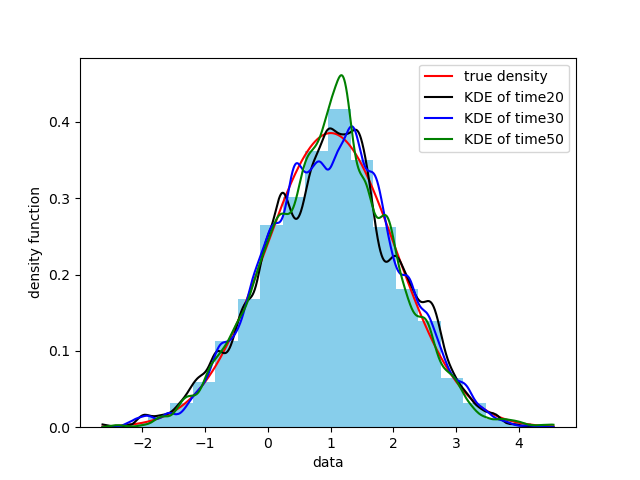}
}
 \subfigure[KDE of F1C2]{
  \centering
  \includegraphics[width=0.22\linewidth,height=3.5cm]{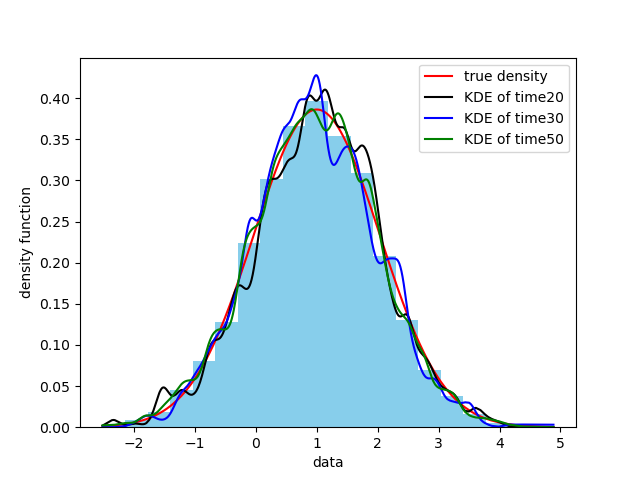}
}
 \subfigure[KDE of F2C1]{
  \centering
  \includegraphics[width=0.22\linewidth,height=3.5cm]{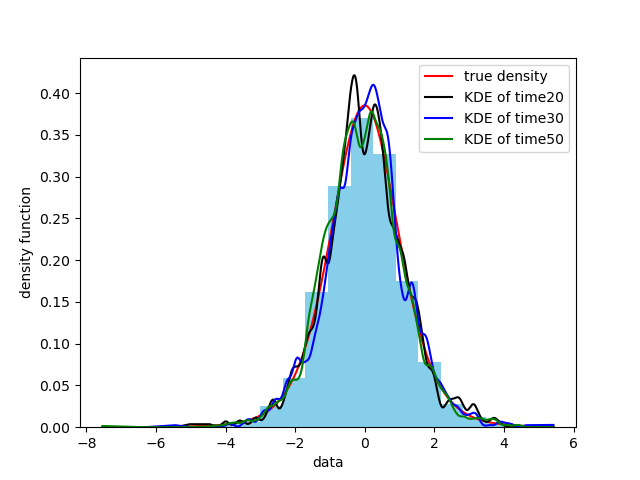}
}
 \subfigure[KDE of F2C2]{
  \centering
  \includegraphics[width=0.22\linewidth,height=3.5cm]{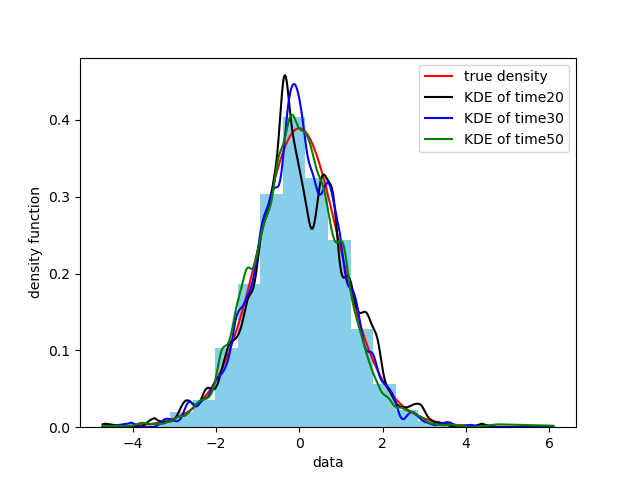}
}
\\
\subfigure[KDE of F1C1]{
  \centering
  \includegraphics[width=0.22\linewidth,height=3.5cm]{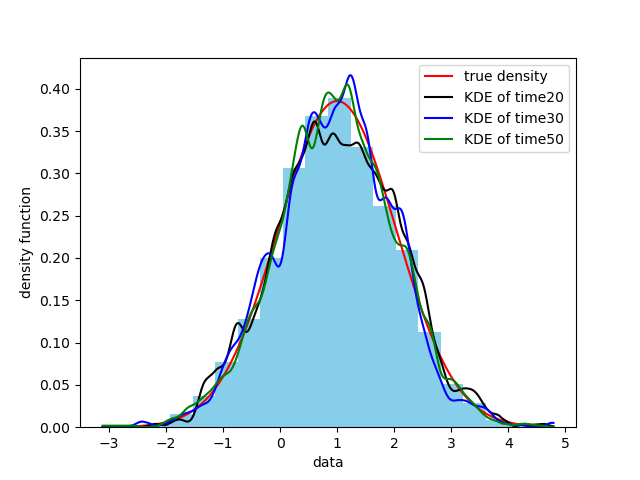}
}
\subfigure[KDE of F1C2]{
  \centering
  \includegraphics[width=0.22\linewidth,height=3.5cm]{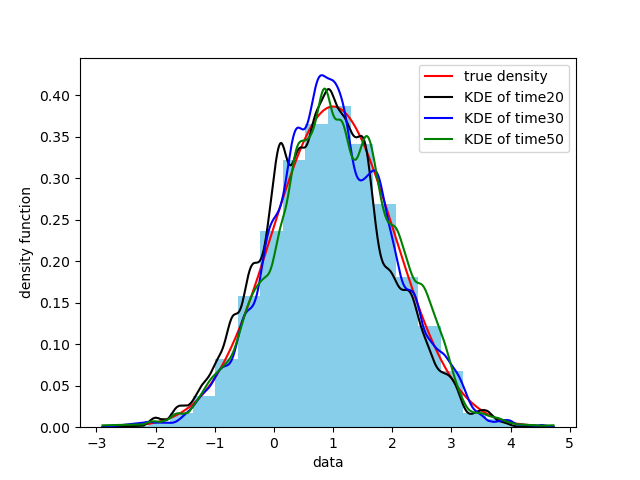}
}
\subfigure[KDE of F2C1]{
  \centering
  \includegraphics[width=0.22\linewidth,height=3.5cm]{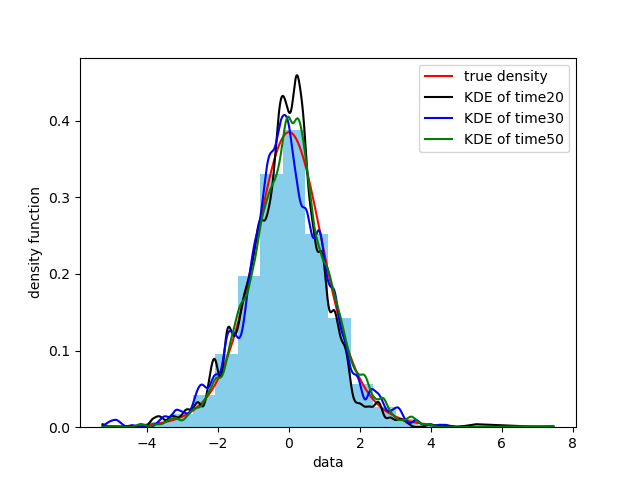}
}
\subfigure[KDE of F2C2]{
  \centering
  \includegraphics[width=0.22\linewidth,height=3.5cm]{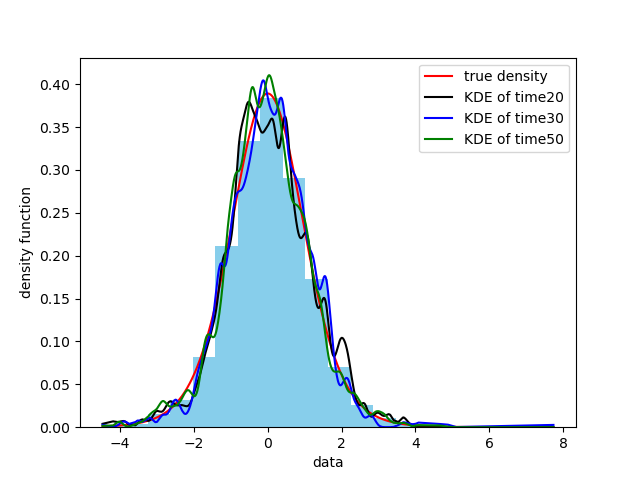}
} 
\caption{KDEs of time lengths 20, 30 and 50 for both features and clusters based on scenarios S1 and S3.  Subfigures (a)$\sim$(d) correspond to S3 with ARIs $\approx$ 0.96.  Subfigures (e)$\sim$(h) correspond to S1 with ARIs $\approx$ 0.54. The black solid line represents the true distribution with other colors representing KDEs with varying time lengths. The blue shadow area is the histogram of sample.}
\label{fig:kernel}
\end{figure}

\begin{figure}[!htp]
  \centering
\subfigure[MSE of feature 1 and cluster 1]{
  \centering
  \includegraphics[width=0.45\linewidth,height= 0.1\textheight]{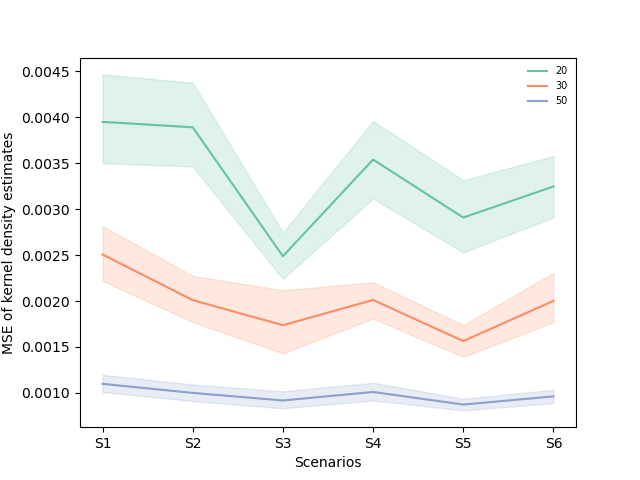}
}
\subfigure[MSE of feature 1 and cluster 2]{
  \centering
  \includegraphics[width=0.45\linewidth,height= 0.1\textheight]{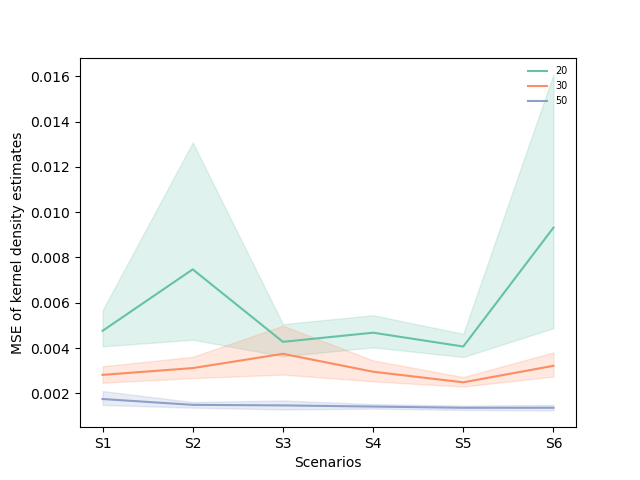}
}\\
\subfigure[MSE of feature 2 and cluster 1]{
  \centering
  \includegraphics[width=0.45\linewidth,height= 0.1\textheight]{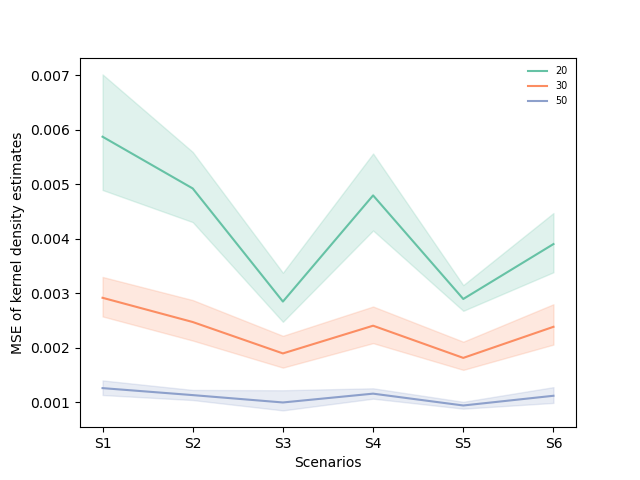}
}
\subfigure[MSE of feature 2 and cluster 2]{
  \centering
  \includegraphics[width=0.45\linewidth,height= 0.1\textheight]{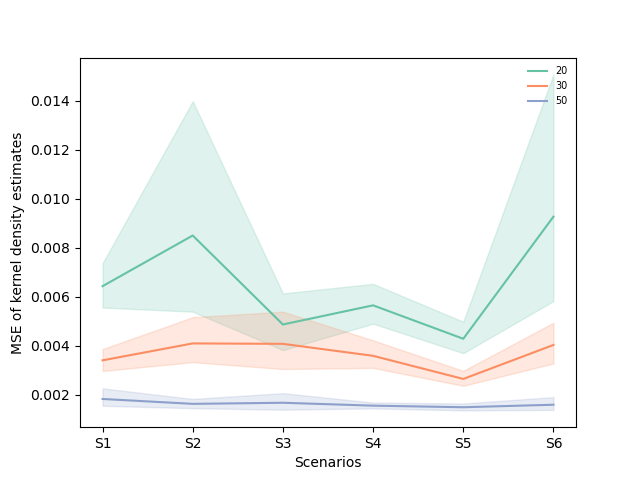}}
\caption{MSE of the KDE for both features and clusters. MSE of the KDE is computed on 100 datasets for each scenario and time length. Colors represent different time lengths, solid lines are the average MSE of 100 datasets over six scenarios, and shadow areas are 95\% confidence interval. }
\label{fig:MSEkernel}
\end{figure}

Figure~\ref{fig:kernel} shows the kernel density estimates (KDEs) of S1 and S3, if bandwidths are known. As can be seen in the Figure, this algorithm tends to choose small bandwidths, yielding undersmoothed estimates. As the KDEs in the CKMM are not used for prediction, under-smoothing is acceptable.  The mean squared error (MSE) of the KDEs is defined by
\begin{align}
    \text{MSE} = \int_{-\infty}^{\infty}(\hat{f}(x)-f(x))^2dx,\nonumber
\end{align}
and Figure~\ref{fig:MSEkernel} shows the calculation results. Note that the MSE tends to decrease as time length increases, and the MSE of feature 1 is smaller than that of feature 2. 

\begin{figure}[!htp]
\centering
\subfigure[MSE of correlation matrix in cluster1]{
  \centering
  \includegraphics[width=0.45\linewidth,height= 4.5cm]{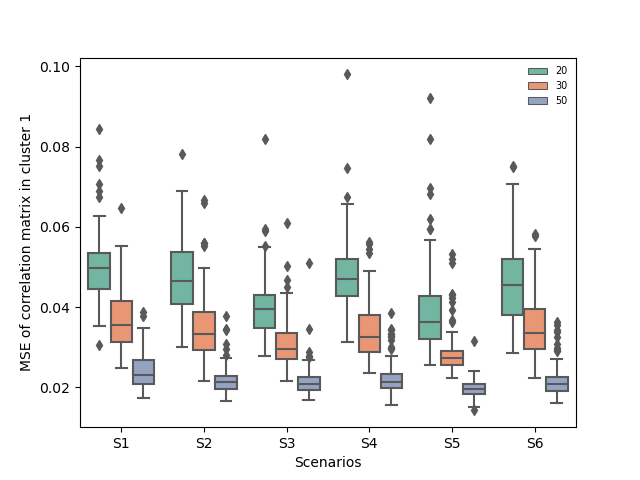}
}
\subfigure[MSE of correlation matrix in cluster2]{
  \centering
  \includegraphics[width=0.45\linewidth,height= 4.5cm]{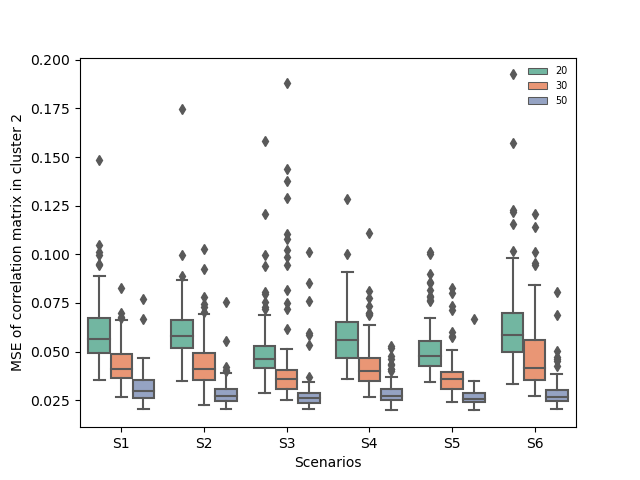}
}
\caption{Boxplots show the MSE of correlations in both clusters. The MSE is computed on 100 datasets for each scenario and time length. Colors represent different time lengths. There are some outliers corresponding to poor performance on certain datasets.}
\label{fig:MSER}
\end{figure}
\begin{figure}[!htp]
\centering
\subfigure[Feature 1 cluster 1]{
  \centering
  \includegraphics[width=0.22\linewidth,height= 0.15\textheight]{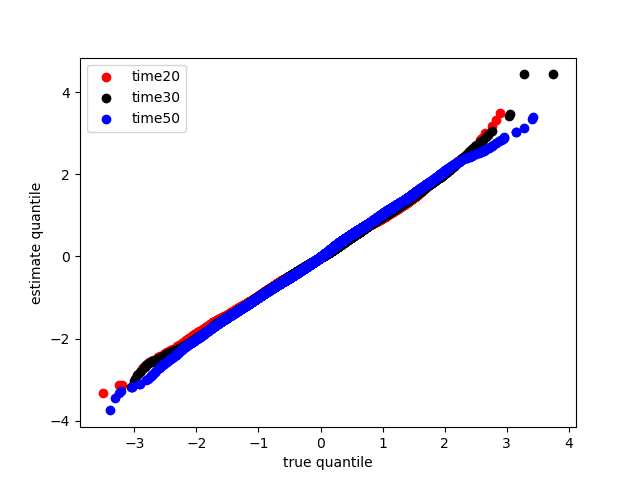}
}
\subfigure[Feature 1 cluster 2]{
  \centering
  \includegraphics[width=0.22\linewidth,height= 0.15\textheight]{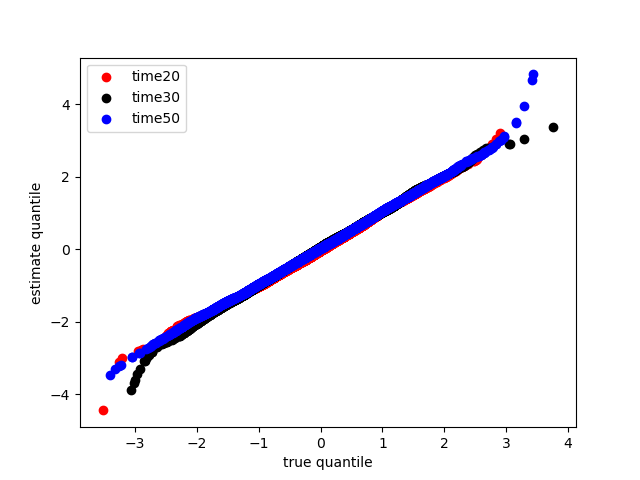}
}
\subfigure[Feature 2 cluster 1]{
  \centering
  \includegraphics[width=0.22\linewidth,height= 0.15\textheight]{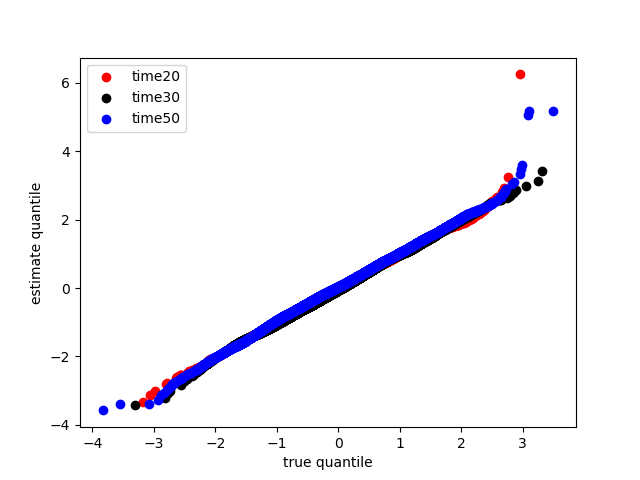}
}
\subfigure[Feature 2 cluster 2]{
  \centering
  \includegraphics[width=0.23\linewidth,height= 0.15\textheight]{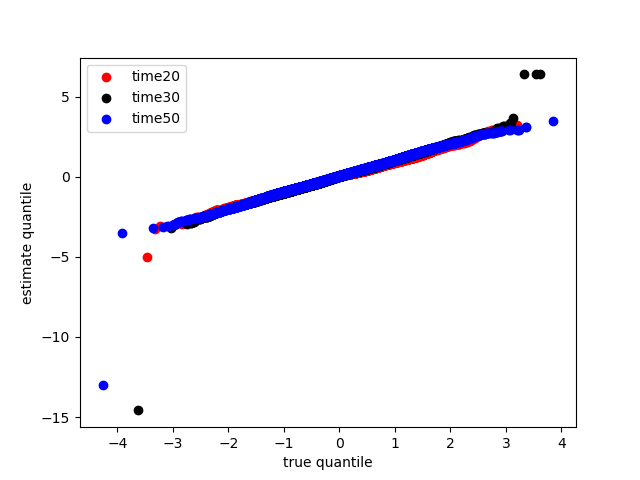}
}\\
\subfigure[Feature 1 cluster 1]{
  \centering
  \includegraphics[width=0.22\linewidth,height= 0.15\textheight]{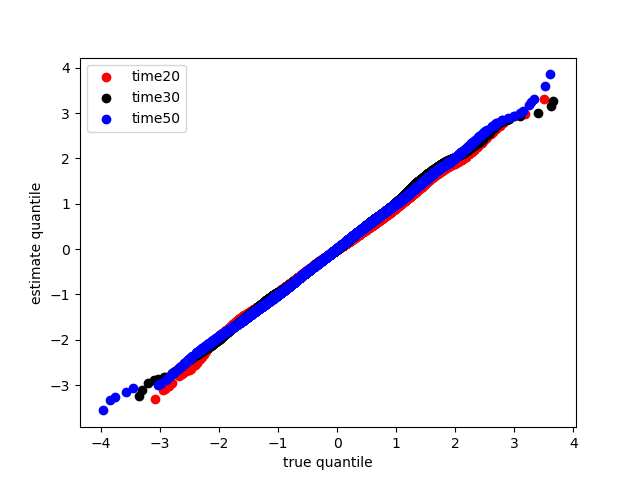}
}
\subfigure[Feature 1 cluster 2]{
  \centering
  \includegraphics[width=0.22\linewidth,height= 0.15\textheight]{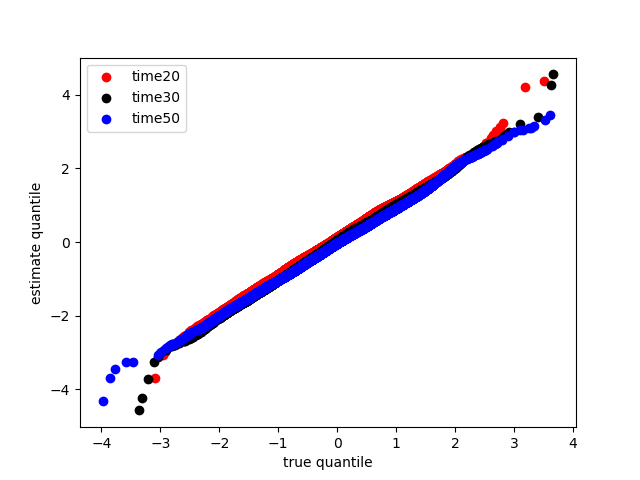}
}
\subfigure[Feature 2 cluster 1]{
  \centering
  \includegraphics[width=0.22\linewidth,height= 0.15\textheight]{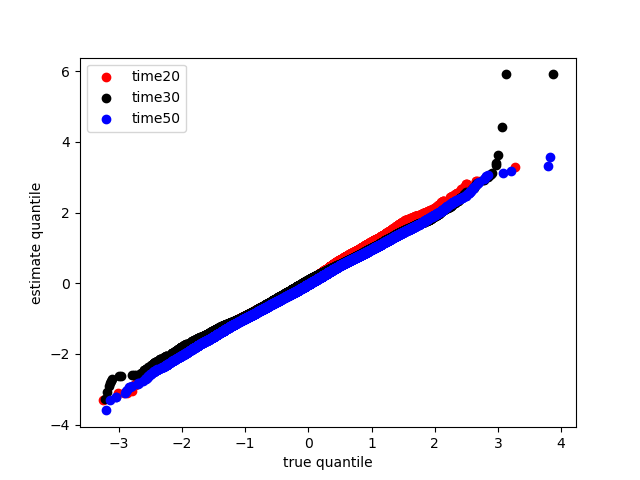}
}
\subfigure[Feature 2 cluster 2]{
  \centering
  \includegraphics[width=0.22\linewidth,height= 0.15\textheight]{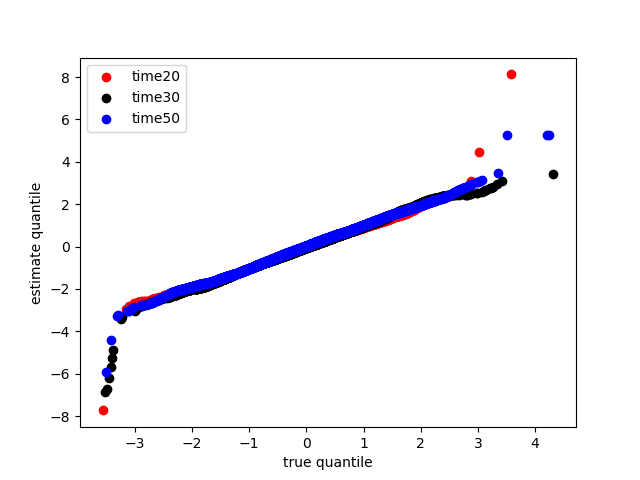}
}
\caption{QQ plots of time lengths 20, 30 and 50 for both features and clusters based on scenarios S1 and S3. The datasets used in this figure are the same as the datasets used in Figure~\ref{fig:kernel}. Red, black, and blue lines represent time lengths 20, 30 and 50, respectively. The top and bottom panels are from S3 and S1, respectively. }
\label{fig:qqplot}
\end{figure}

The MSE of the correlation estimates are displayed in Figure~\ref{fig:MSER}. Like for the KDEs, the MSE tends to decrease as time length increases. It is worth noting that the MSE of S3 is the smallest among the six scenarios, even though the cross-correlations are the largest among the first three scenarios.  In addition, the MSE from cluster 2 is larger than that of cluster 1, which appears to be due to the heaviness of the tail of the t distribution for cluster 2 which may yield some extreme values. The results appear to be influenced by these extreme values, as can be seen in Figure~\ref{fig:qqplot}. Meanwhile, the bias decreases as time length increases, as the bias between circulant and Toplitz matrices decreases as time length increases. 

\subsection{Application}
In this section, the CKMM is used for clustering two real datasets: the Epilepsy and RacketSports datasets, which are available from the Time Series Classification Repository \citep{Bagnall18}. Both datasets have a balanced longitudinal structure and only the training data are used for the clustering.  
\subsubsection{Epilepsy Data}
The Epilepsy training dataset consists of four different activities performed by 3 healthy participants, where each participant performs each activity multiple times. These activities are epilepsy, walking, running and sawing ($G = 4$). The aim of this analysis is to cluster these four activities. The number of repetitions of each activity are 34, 34, 36 and 30 ($N = 134$), and these activities are recorded by a tri-axial accelerometer ($D = 3$). Each feature has 206 observations ($T=206$).
  
\begin{figure}[!htp]
  \centering
\subfigure[Feature 1]{
\centering
\includegraphics[width=0.3\linewidth,height = 0.2\textheight]{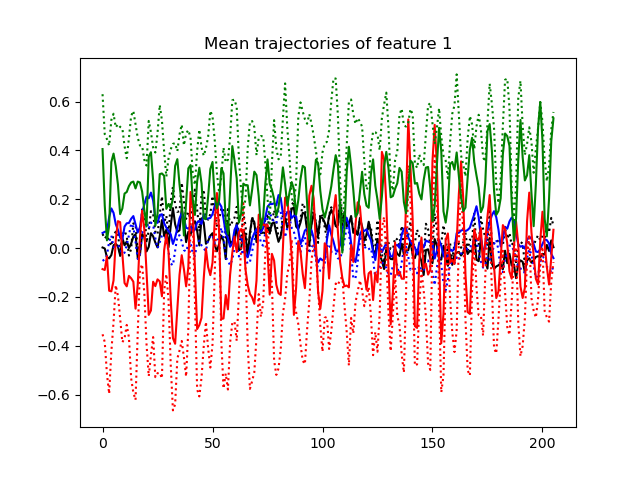}  
}
\subfigure[Feature 2]{
\centering
\includegraphics[width=0.3\linewidth,height = 0.2\textheight]{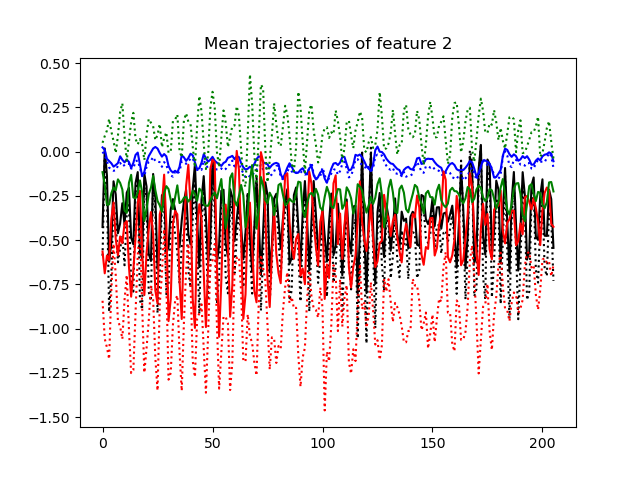}  
}
\subfigure[Feature 3]{
\centering
\includegraphics[width=0.3\linewidth,height = 0.2\textheight]{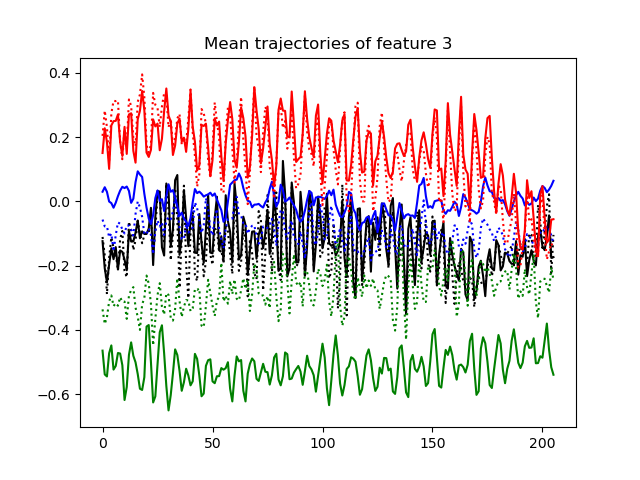}  
}
\caption{Mean trajectories in true and predicted groups. Black, blue, red and green lines represent epilepsy, walking, running and sawing, respectively. True and predicted labels are presented by solid and dotted lines.  The amplitudes of fluctuation in the first two groups are significantly smaller than those in other two groups. }
\label{Fig:app1}
\end{figure}

The mean trajectories of four activities are shown in Figure~\ref{Fig:app1} with solid lines. The amplitudes of fluctuation of epilepsy and walking groups are significantly smaller than those of running and sawing groups. There are no obvious trends in three features, so we assume that series are stationary.  As the CKMM may converge to different solutions, ten different initializations of labels are used. The adjusted BIC and NEC are used to select the number of clusters. The performance of CKMM is compared with DTW and GMM. The coefficients in growth model may have random effects, so LCGA is replaced by GMM. The results are recorded in Table~\ref{tab:BICreal}.

\begin{table}[!htp]
\caption{Adjusted BIC scores, normalized entropy from CKMM and ARI comparison from CKMM, DTW and GMM based on the Epilepsy data set.
\label{tab:BICreal}}
\centering\begin{tabular*}{0.8\textwidth}{@{\extracolsep{\fill}}lcc|lcc}
\hline
\multirow{2}{*}{clusters} & \multirow{2}{*}{Adjusted BIC} & \multirow{2}{*}{NEC} & \multicolumn{3}{c}{ARI} \\ \cline{4-6} 
                          &                      &                      & CKMM        & DTW & GMM \\ \hline
1        & 65587.5840  & 3.3698e-11  & --&--&-- \\
2        & 38204.7654    & \textbf{1.7344e-28}       &0.3247&0.2855&0.1341   \\
3        & 35063.6920     & 2.5947e-21    & 0.4484 &0.4166&0.0841\\
4        & \textbf{27691.8172}    & 1.2633e-12    &\textbf{0.6020}&0.4225&0.0885\\
5        & 30498.0215     & 9.0370e-09      & 0.4954&0.3349&0.1385 \\ 
 \hline
\end{tabular*}
\end{table}

As shown in Table~\ref{tab:BICreal}, the NEC selects $G=2$ and the adjusted BIC selects $G=4$. The CKMM has the highest ARI for each fixed number of clusters compared to the other methods, with the highest ARI of 0.6020 selected by the adjusted BIC at $G=4$. The confusion matrix of $G=2$ and $G=4$ are shown in appendix~\ref{appxB}, Tables~\ref{tab:confusionk2} and \ref{tab:confusionk4}. 

 When the number of clusters is 4, the predicted mean trajectories of three features are shown in Figure~\ref{Fig:app1} with dotted lines. Note that the predicted mean trajectories of sawing and running groups are far away from the true mean trajectories, as 9 running cases with high amplitudes are misclassified to sawing groups. 

\subsubsection{RacketSports Data}
The RacketSports training data set was generated by 151 students playing squash or badminton. There are two different kinds of strokes in each sport: forehand/backhand in squash and clears/smashes in
badminton. The aim of this data set is to identify which sport and which stroke the players are making. Six features are collected from a smart watch and each feature has 30 observations.

As for the previous data set, ten different initializations are used for the CKMM. The Adjusted BIC, NEC and ARIs are shown in Table~\ref{tab:Racket}. The lowest adjusted BIC happens at $K=2$, but NEC selects $K=5$. The ARIs of the CKMM are the highest amongst the three clustering methods.  The confusion matrices of $K=2$ and $K=5$ are shown in Appendix~\ref{appxB}, Tables~\ref{tab:Rconfusionk2} and \ref{tab:Rconfusionk3}. Table~\ref{tab:Rconfusionk4} in Appendix~\ref{appxB} is the confusion matrix if the number of clusters is known.

\begin{figure}[!htp]
  \centering
\subfigure[Feature 1]{
\centering
\includegraphics[width=0.3\linewidth,height = 0.15\textheight]{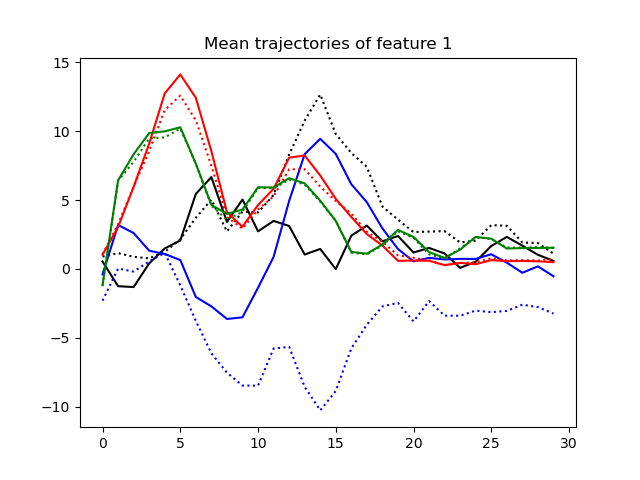}  
}
\subfigure[Feature 2]{
\centering
\includegraphics[width=0.3\linewidth,height = 0.15\textheight]{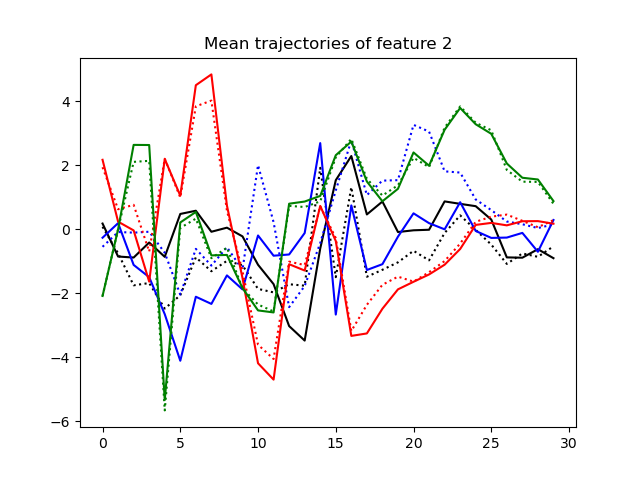}  
}
\subfigure[Feature 3]{
\centering
\includegraphics[width=0.3\linewidth,height = 0.15\textheight]{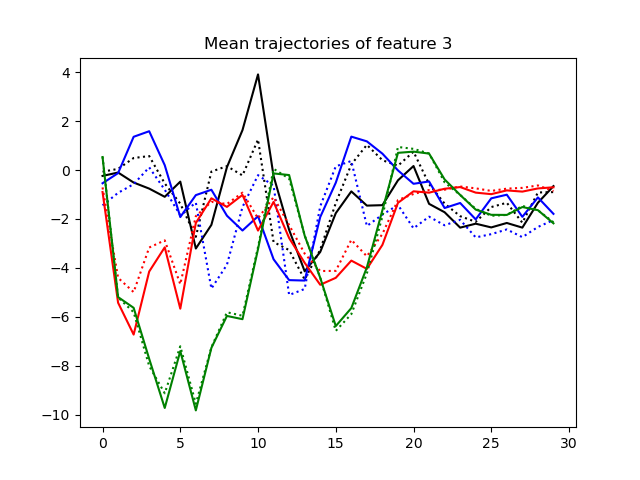}  
}\\
\subfigure[Feature 4]{
\centering
\includegraphics[width=0.3\linewidth,height = 0.15\textheight]{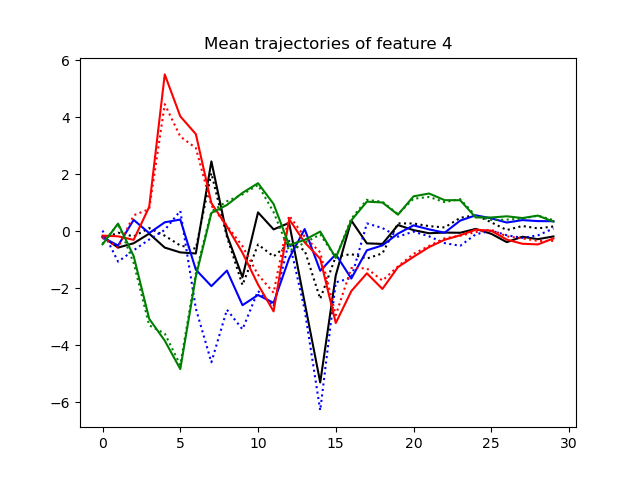}  
}
\subfigure[Feature 5]{
\centering
\includegraphics[width=0.3\linewidth,height = 0.15\textheight]{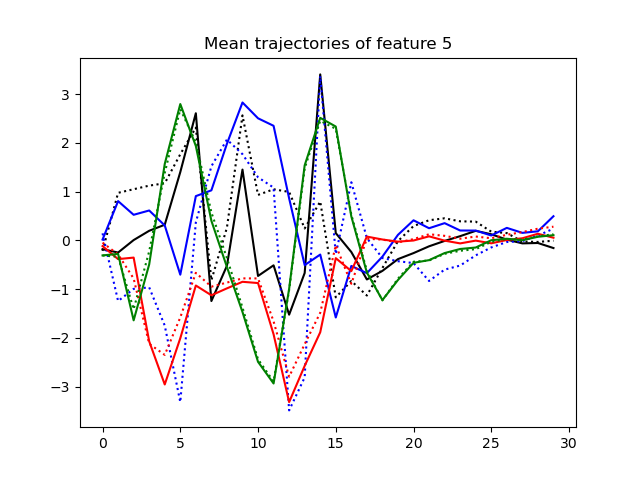}  
}
\subfigure[Feature 6]{
\centering
\includegraphics[width=0.3\linewidth,height = 0.15\textheight]{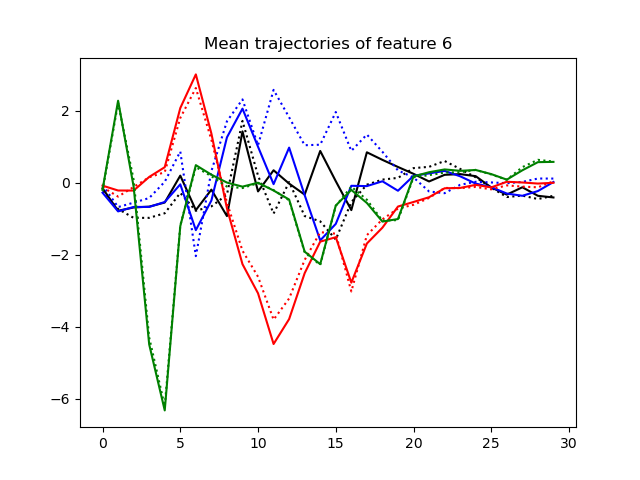}  
}
\caption{Mean trajectories in true and predicted groups for the Racketsports training dataset. Black, blue, red, and green lines represent smash, clear, forehand and backhand respectively. The solid and dotted lines are from true and predicted labels.}
\label{Fig:app2}
\end{figure}

\begin{table}[!htp]
\caption{Adjusted BIC and normalized entropy for the CKMM and ARI comparison between the CKMM, DTW and GMM on the RacketSports data set.
\label{tab:Racket}}
\centering\begin{tabular*}{0.8\textwidth}{@{\extracolsep{\fill}}lcc|lcc}
\hline
\multirow{2}{*}{clusters} & \multirow{2}{*}{Adjusted BIC} & \multirow{2}{*}{NEC} & \multicolumn{3}{c}{ARI} \\ \cline{4-6} 
                          &                      &                      & CKMM        & DTW & GMM \\ \hline
1        & 148228.9962  & 2.2335e-04  & --&--&-- \\
2        & \textbf{148086.8007}   & 6.5547e-05    &0.4427 &0.4413 &0.0733   \\
3        & 149689.1247     & 2.6539e-06  & 0.3494 &0.3276 &0.1173\\
4        & 151439.6019   & 4.9985e-06&0.5484&0.3410&0.1425\\
5        & 154488.8859   & \textbf{4.6486e-09}&\textbf{0.5821}&0.4538&0.1389\\
 \hline
\end{tabular*}
\end{table}

Tables~\ref{tab:Rconfusionk2}$\sim$\ref{tab:Rconfusionk4} show it is easier to identify the type of sport, but the stroke types in badminton are difficult to distinguish. Figure~\ref{Fig:app2} shows the mean trajectories in the true and predicted groups. As the number of misclassified strokes in badminton is large, predicted mean trajectories of smash and clear deviate from the true trajectories. However, predicted trajectories for squash are close to the true curves.

\section{Discussion}
\label{sec5}
The CKMM is proposed as a copula-based approach for clustering multivariate longitudinal data. The copula framework decomposes a joint distribution into its margins and copula. This framework gives a mechanism for forming more flexible, valid joint distributions through different specifications of valid marginal and copula distributions. We propose a semi-parametric approach, where the copula is from the Gaussian parametric family, but the marginal density functions are estimated by a kernel functions, rendering marginal parametric assumptions unnecessary. Furthermore, the circulant matrix is used to approximate correlation matrix of the Gaussian copula, leading to a significant decrease in the number of free parameters. 

The performance of the proposed method was investigated in a simulation study. The results showed that the MSE of both the kernel density and correlation estimates decrease as time length increases. As the kernel function is the Gaussian kernel, the kernel density estimates fit the Gaussian distribution better than the t distribution, especially when the time length is short. The results also showed the MSE of the correlation matrices decreases as cross-correlation increases. 

The performances of $K$-means with DWT, longitudinal latent growth models, and CKMM were compared using ARI in a simulation study and data applications. In the simulation study, the CKMM was shown to outperform the other two methods across all simulated scenarios. The performance of the LCGA was very poor when correlations were similar between clusters, but improved as the difference between correlations increased. $K$-means with DTW distance had an overall poor performance in the simulation study. In the real data analysis, the ARIs of the CKMM were the highest among the three methods for each fixed number of clusters. In the data analyses, $K$-means with DTW performed better than the longitudinal latent growth model (the GMM). 

Although the CKMM performs reasonably well for balanced longitudinal data, there is still room for improvement. The purpose of this work was to propose a new method for clustering multivariate longitudinal data that uses marginal, multivariate dependence, and time dependence information. However, as with most model-based methods, it relies on model selection. Traditional information criteria use penalty terms based on Euclidean parameters and are not tailored to functional parameters. Hence, one area of potential improvement is develop a new criterion for better selection of the number of clusters and copula distribution. In addition, the CKMM is unable to handle longitudinal data with unequal time lengths. Such data can appear often in practice and therefore future work will focus on extensions of the CKMM to unbalanced longitudinal data.

\appendix
\section{Mathematical Proof}
\label{appxA}
\subsection{Constraint on Correlation Matrices}
\label{appx:correlation}
The autocovariance of $X_{dt}$ and $X_{ds}$ is
\begin{align}
    \text{cov}(X_{dt},X_{ds})\!= E[F_{dt}^{-1}\{\Phi(q_{dt})\}F_{ds}^{-1}\{\Phi(q_{ds})\}]-E[F_{dt}^{-1}\{\Phi(q_{dt})\}]E[F_{ds}^{-1}\{\Phi(q_{ds})\}].\nonumber
\end{align}
Next, we will prove that if the correlation matrix $\vecR$ in the copula function depends solely on the time lags, then the auto-covariance of the time series is also determined by the time lags, indicating that the time series is stationary. Hence, imposing a constraint on $\vecR$ can satisfy the stationary assumption. Using a Taylor expansion to estimate above covariance yields
\begin{align}
    &E[F_{dt}^{-1}\{\Phi(q_{dt})\}F_{ds}^{-1}\{\Phi(q_{ds})\}]\nonumber\\
    &\approx E\left(F_{dt}^{-1}(0.5)F_{ds}^{-1}(0.5)+q_{dt}\frac{1}{\sqrt{2\pi}f_{dt}\{F_{dt}^{-1}(0.5)\}}F_{ds}^{-1}(0.5)+q_{ds}\frac{1}{\sqrt{2\pi}f_{ds}\{F_{ds}^{-1}(0.5)\}}F_{dt}^{-1}(0.5)\right.\nonumber\\
    &\quad\left.+\frac{1}{2}\left[-q^2_{dt}\frac{f^{'}\{F_{dt}^{-1}(0.5)\}}{{2\pi}f^3_{dt}\{F_{dt}^{-1}(0.5)\}}F_{ds}^{-1}(0.5)+2q_{dt}q_{ds}\frac{1}{\sqrt{2\pi}f_{dt}\{F_{dt}^{-1}(0.5)\}}\frac{1}{\sqrt{2\pi}f_{ds}\{F_{ds}^{-1}(0.5)\}}\right.\right.\nonumber\\
    &\quad\left.\left.-q^2_{ds}\frac{f^{'}\{F_{ds}^{-1}(0.5)\}}{{2\pi}f^3_{ds}\{F_{ds}^{-1}(0.5)\}}F_{dt}^{-1}(0.5)\right]\right)\nonumber\\
    &\approx F_{dt}^{-1}(0.5)F_{ds}^{-1}(0.5)+\frac{1}{\sqrt{2\pi}f_{dt}\{F_{dt}^{-1}(0.5)\}}\frac{1}{\sqrt{2\pi}f_{ds}\{F_{ds}^{-1}(0.5)\}}E(q_{dt}q_{ds})\nonumber\\
    &\quad-\frac{F_{ds}^{-1}(0.5)f^{'}\{F_{dt}^{-1}(0.5)\}}{{4\pi}f^3_{dt}\{F_{dt}^{-1}(0.5)\}}E(q^2_{dt})-\frac{F_{dt}^{-1}(0.5)f^{'}\{F_{ds}^{-1}(0.5)\}}{{4\pi}f^3_{ds}\{F_{ds}^{-1}(0.5)\}}E(q^2_{ds})\nonumber\\
    &E[F_{dt}^{-1}\{\Phi(q_{dt})\}]\approx F_{dt}^{-1}(0.5)-\frac{f^{'}\{F_{dt}^{-1}(0.5)\}}{{4\pi}f^3_{dt}\{F_{dt}^{-1}(0.5)\}}E(q^2_{dt})\nonumber\\
    &E[F_{ds}^{-1}\{\Phi(q_{ds})\}]\approx F_{ds}^{-1}(0.5)-\frac{f^{'}\{F_{ds}^{-1}(0.5)\}}{{4\pi}f^3_{ds}\{F_{ds}^{-1}(0.5)\}}E(q^2_{ds})\nonumber.
\end{align}
Then the covariance of $X_{dt}$ and $X_{ds}$ is approximated by
\begin{align}
  \text{cov}(X_{dt},X_{ds}) \approx   \frac{1}{{2\pi}f_{dt}\{F_{dt}^{-1}(0.5)\}f_{ds}\{F_{ds}^{-1}(0.5)\}}\text{cov}(q_{dt},q_{ds})\!+\!\frac{f^{'}\{F_{dt}^{-1}(0.5)\}f^{'}\{F_{ds}^{-1}(0.5)\}}{{16\pi^2}f^3_{dt}\{F_{dt}^{-1}(0.5)\}f^3_{ds}\{F_{ds}^{-1}(0.5)\}}\nonumber.
\end{align}
Hence, if $\text{cov}(q_{dt},q_{ds})$ only depends on the time lags, so too does $\text{cov}(X_{dt},X_{ds})$.
\subsection{Identifiability}
Theorem 2 proposed by \cite{teicher1963identifiability} and 
 Corollary 2.1 proposed by \cite{chandra1977mixtures} will be used to prove the identifiability of CKMM. Let $\mathcal{F}=\left\{F\right\}$ be a family of component density functions with a transforms $\phi(t)$ defined for $t \in \mathcal{S}_{\phi}$.  Theorem 2 from \cite{teicher1963identifiability} proved that if there exists a total ordering denoted by $F_{1} \preceq F_{2}$, which satisfies: (i) $\mathcal{S}_{\phi_1} \subseteq  \mathcal{S}_{\phi_2}$ and (ii) $\exists t_{1} \in  \mathcal{S}_{\phi_1}$, $\lim_{t \to t_{1}} \phi_2(t)/\phi_1(t)=0$, then the finite mixture model is identifiable. Corollary 2.1 from \cite{chandra1977mixtures} proved that if the mixing distributions relative to marginal distribution are identifiable, then the mixing distributions of the corresponding joint distribution is identifiable. 
 
 Two finite subsets from $\mathcal{F}$ are denoted by $\mathcal{F}_{1}=\{c(\mathbf{q}\mid\mathbf{R}_{g})\prod_{d=1}^{D}\prod _{t=1}^{T}f_{d,g}(x_{dt}),1\leq g\leq G\}$ and $\mathcal{F}_{2}=\{c(\hat{\mathbf{q}}\mid\hat{\mathbf{R}}_{g})\prod_{d=1}^{D}\prod_{t=1}^{T}\hat{f}_{d,g}(x_{dt}),1\leq g\leq \hat{G}\}$. 
 If the copula kernel mixture model is identifiable, then 
 \begin{align}
     \sum_{g=1}^{G}\pi_{g}\left\{c(\mathbf{q}\mid\mathbf{R}_{g})\prod_{d=1}^{D}\prod_{t=1}^{T}f_{d,g}(x_{dt})\right\}\equiv\sum_{g=1}^{\hat{G}}\hat{\pi}_{g}\left\{c(\hat{\mathbf{q}}\mid\hat{\mathbf{R}}_{g})\prod_{d=1}^{D}\prod_{t=1}^{T}\hat{f}_{d,g}(x_{dt})\right\}\nonumber
 \end{align}
 implies that $G=\hat{G}$, $\pi_{g}=\hat{\pi}_g$, $\mathbf{R}_{g}=\hat{\mathbf{R}}_{g}$, and $f_{d,g}(x_{dt})=\hat{f}_{d,g}(x_{dt})$. The mixing distribution of corresponding marginals will be :
 \begin{align}
     \sum_{g=1}^{G}\pi_{g}\left\{\sum_{n=1}^{N}\sum_{j=1}^{T}\frac{p_{ng}}{T\sum_{n=1}^{T}p_{ng}}K_{h_{d,g}}(x_{ndj},x_{dt})\right\} \!=\! \sum_{k=1}^{\hat{G}}\hat{\pi}_{k}\left\{\!\sum_{n=1}^{N}\sum_{j=1}^{T}\frac{\hat{p}_{nk}}{T\sum_{n=1}^{T}\hat{p}_{nk}}K_{\hat{h}_{d,k}}(x_{ndk},x_{dt})\!\right\}.\nonumber
 \end{align}
 By corollary 2.1 of \cite{chandra1977mixtures}, if mixing distribution of marginals identifiable, then CKMM will be identifiable. Next, theorem 2 from \cite{teicher1963identifiability} will be applied to show the identifiability of mixing distribution of marginals.
 
 Data points $x_{ndt}$ and bandwidth $h_{d,g}$ are parameters in the kernel functions.  Suppose that the total ordering introduced in theorem 2 exists in kernel functions via reordering parameters. The ordering is $K_{h_{d,(1)}}(x_{(1d1)},x_{dt})\preceq \cdots\preceq K_{h_{d,(1)}}(x_{(NdT)},x_{dt})\preceq K_{h_{d,(2)}}(x_{(1d1)},x_{dt})\preceq \cdots\preceq K_{h_{d,(G)}}(x_{(NdT)},x_{dt})$ , {where $h_{d,(g)}$ and $x_{(ndt)}$ for $ n =1,\dots,N; g=1,\dots,G; t=1,\dots,T$ are reordered parameters}  and transform function is denoted by $\phi(s\mid x_{(ndj)},h_{d,(g)})$. The kernel functions with $\hat{h}$ also follow this rule, and above equation can be written as
  \begin{align}
&\sum_{g=1}^{G}\pi_{(g)}\left\{\sum_{n=1}^{N}\sum_{j=1}^{T}\frac{p_{(ng)}}{T\sum_{n=1}^{N}p_{(ng)}}\phi (s\mid x_{(ndj)},h_{d,(g)})\right\} \nonumber\\
=&\! \sum_{k=1}^{\hat{G}}\hat{\pi}_{(k)}\left\{\!\sum_{n=1}^{N}\sum_{j=1}^{T}\frac{\hat{p}_{(nk)}}{T\sum_{n=1}^{N}\hat{p}_{(nk)}}\phi (s\mid x_{(ndj)},\hat{h}_{d,(g)})\!\right\}.\nonumber
 \end{align}
If $K_{h_{d,(1)}}(x_{(1d1)},x_{dt}) \neq K_{\hat{h}_{d,(1)}}(x_{(1d1)},x_{dt})$, and  assume $K_{h_{d,(1)}}(x_{(1d1)},x_{dt}) \preceq K_{\hat{h}_{d,(1)}}(x_{(1d1)},x_{dt})$, then dividing $\phi(s\mid x_{(1d1)},h_{d,(1)})$ both sides yields
\begin{align}
    &\frac{\pi_{(1)}p_{(11)}}{T\sum_{n}p_{(n1)}}\!+\!\mathop{\sum\sum\sum}_{x_{(ndj)}\neq x_{(1d1)}} \frac{\pi_{(g)}p_{(ng)}}{T\sum_{n}p_{(ng)}}\frac{\phi(s\mid x_{(ndj)},h_{d,(g)})}{\phi(s\mid x_{(1d1)},h_{d,(1)})}\nonumber\\
    &= \! \sum_{k=1}^{\hat{G}}\sum_{n=1}^{N}\sum_{j=1}^{T}\frac{\hat{\pi}_{(k)}\hat{p}_{(nk)}}{T\sum_{n=1}^{T}\hat{p}_{(nk)}}\frac{\phi (s\mid x_{(ndj)},\hat{h}_{d,(k)})}{\phi (s\mid x_{(ndj)},h_{d,(1)})}.\nonumber
\end{align}
Given that
\begin{align}
   & \lim_{s \to s_{1}} \frac{\phi (s\mid x_{(ndj)},h_{d,(g)})}{\phi (s\mid x_{(1d1)},h_{d,(1)})}=0 \qquad \lim_{s \to s_{1}} \frac{\phi (s\mid x_{(ndj)},\hat{h}_{d,(k)})}{\phi (s\mid x_{(1d1)},h_{d,(1)})}=0, \nonumber\\
   &\implies \frac{\pi_{(1)}p_{(11)}}{T\sum_{n}p_{(n1)}} =0.\nonumber
\end{align}
However, $\pi_{(1)}$ and $p_{(11)}$ are positive, so the assumption does not hold and $K_{h_{d,(1)}}(x_{(1d1)},x_{dt}) = K_{\hat{h}_{d,(1)}}(x_{(1d1)},x_{dt})$, which implies $h_{d,(1)} = \hat{h}_{d,(1)}$. Meanwhile, we get 
\begin{align}
    &\frac{\pi_{(1)} p_{(11)} }{\sum_{n}p_{(n1)}} = \frac{\hat{\pi}_{(1)} \hat{p}_{(11)} }{\sum_{n}\hat{p}_{(n1)}} \nonumber\\
    \implies &\frac{\hat{p}_{(11)}}{p_{(11)}} = \frac{\pi_{(1)}}{\hat{\pi}_{(1)}}\frac{\sum_{n}\hat{p}_{(n1)}}{\sum_{n}{p}_{(n1)}}.\nonumber
\end{align}
This ratio does not vary with subjects, so $\pi_{(1)} = \hat{\pi}_{(1)}$ and $p_{(11)} = \hat{p}_{(11)}$. The proof of $G = \hat{G}$ is the same as  the proof in \cite{teicher1963identifiability}. Thus, the mixing distributions of marginals  are identifiable as long as kernel functions satisfy the total ordering in theorem 2 from \cite{teicher1963identifiability}. 

The kernel function {used in this paper} is a Gaussian kernel, and the corresponding one-to-one mapping is moment generating function:
\begin{align}
    \phi(x_{dt}\mid x_{(ndj)},h_{d,g})=\exp\left\{\frac{h^2_{d,g}}{2}s^2-x_{(ndj)}s\right\}.\nonumber
\end{align}
Next, we will prove that the total ordering exists in  Gaussian kernel functions if bandwidths vary with clusters and data points used in kernels are different. Reordering bandwidths makes them satisfy that $h^2_{d,(1)}>\cdots>h^2_{d,(G)}$ , and then
\begin{align}
     &\lim_{s \to \infty}\frac{\phi(x_{dt}\mid x_{(ndj)},h_{d,(g)})}{\phi(x_{dt}\mid x_{(mdk)},h_{d,(1)})}= \lim_{s \to \infty}\exp\left\{\frac{h^2_{d,(g)}-h^2_{d,(1)}}{2}s^2-\left(x_{(ndj)}-x_{(mdk)}\right)s\right\}=0.\nonumber
\end{align}
Reordering data points $x_{ndt}$ for $n=1,\dots,N$ and $t=1,\cdots,T$ makes them satisfy that $x_{(1d1)}<\cdots<x_{(NdT)}$, and then 
\begin{align}
     &\lim_{s \to \infty}\frac{\phi(x_{dt}\mid x_{(ndj)},h_{d,(1)})}{\phi(x_{dt}\mid x_{(1d1)},h_{d,1})}=\exp\left\{-\left(x_{(ndj)}-x_{(1d1)}\right)s\right\}=0.\nonumber
\end{align}
Hence, the total ordering exists in marginal distributions, and mixing distribution of marginals based on Gaussian kernel functions is identifiable. Then, by the corollary 2.1 of \cite{chandra1977mixtures}, the identifiability of CKMM based on Gaussian functions is proved.

\section{Confusion Matrices in Applications}
\label{appxB}
\begin{table}[H]
\caption{Confusion matrix of CKMM when $K$ = 2 on Epilepsy.}
\centering\begin{tabular*}{0.8\textwidth}{@{\extracolsep{\fill}}lcccr}
\hline
   &  epilepsy & walking  & running&sawing  \\\hline
cluster 1 &9  &37 & 0  &22\\
cluster 2 &25   &0   &36  &8\\
\hline
\end{tabular*}
\label{tab:confusionk2}
\end{table}

\begin{table}[H]
\caption{Confusion matrix of CKMM when $K$ = 4 on Epilepsy.}
\centering\begin{tabular*}{0.8\textwidth}{@{\extracolsep{\fill}}lcccr}
\hline
   &  epilepsy & walking  & running&sawing \\\hline
cluster 1 &3  &0  &9  &24  \\
cluster 2 &25 &0  &0  &0 \\
cluster 3 &6  &37 &0  &4\\
cluster 4 &0  &0  &27 &2\\
\hline
\end{tabular*}
\label{tab:confusionk4}
\end{table}

\begin{table}[H]
\caption{Confusion matrix of CKMM when $K$ = 2 on Racketsports.}
\centering\begin{tabular*}{0.8\textwidth}{@{\extracolsep{\fill}}lcccr}
\hline
   &  Smash & Clear  &Forehand&Backhand \\\hline
cluster 1 &38  &40 & 0  &0\\
cluster 2 &1   &3   &35  &34\\
\hline
\end{tabular*}
\label{tab:Rconfusionk2}
\end{table}

\begin{table}[H]
\caption{Confusion matrix of CKMM when $K$ = 5 on Racketsports.}
\centering\begin{tabular*}{0.8\textwidth}{@{\extracolsep{\fill}}lcccr}
\hline
  &  Smash & Clear  &Forehand&Backhand \\\hline
cluster 1 &11  &11  &0  &0  \\
cluster 2 &1 &0  &0  &34 \\
cluster 3 &0  &3 &35  &0\\
cluster 4 &8  &0 &0  &0\\
cluster 5 &19  &29 &0  &0\\
\hline
\end{tabular*}
\label{tab:Rconfusionk3}
\end{table}

\begin{table}[H]
\caption{Confusion matrix of CKMM when $K$ = 4 on Racketsports.}
\centering\begin{tabular*}{0.8\textwidth}{@{\extracolsep{\fill}}lcccr}
\hline
  &  Smash & Clear  &Forehand&Backhand \\\hline
cluster 1 &1  &4  &35  &1  \\
cluster 2 &26 &28  &0  &0 \\
cluster 3 &1  &0 &0  &33\\
cluster 4 &11  &11 &0  &0\\
\hline
\end{tabular*}
\label{tab:Rconfusionk4}
\end{table} 
 \subsection*{Acknowledgements}
This work was supported by respective NSERC Discovery Grants (OM, PM), the Canada Research Chairs program (PM), and a Dorothy Killam Fellowship (PM).

 \subsection*{Declarations}
The authors have no relevant financial or non-financial interests to disclose.
 \subsection*{Data Availability} The Time Series Classification Repository freely provided the data used in this study (\url{http://www.timeseriesclassification.com/}).

{\small
\bibliographystyle{chicago}
\bibliography{Xi}

\begin{thebibliography}{}

\bibitem[\protect\citeauthoryear{Abraham, Cornillon, Matzner-L{\o}ber, and
  Molinari}{Abraham et~al.}{2003}]{abraham2003unsupervised}
Abraham, C., P.-A. Cornillon, E.~Matzner-L{\o}ber, and N.~Molinari (2003).
\newblock Unsupervised curve clustering using b-splines.
\newblock {\em Scandinavian journal of statistics\/}~{\em 30\/}(3), 581--595.

\bibitem[\protect\citeauthoryear{Bagnall, Dau, Lines, Flynn, Large, Bostrom,
  Southam, and Keogh}{Bagnall et~al.}{2018}]{Bagnall18}
Bagnall, A., H.~A. Dau, J.~Lines, M.~Flynn, J.~Large, A.~Bostrom, P.~Southam,
  and E.~Keogh (2018).
\newblock The uea multivariate time series classification archive, 2018.
\newblock {\em arXiv preprint arXiv:1811.00075\/}.

\bibitem[\protect\citeauthoryear{Berndt and Clifford}{Berndt and
  Clifford}{1994}]{berndt1994using}
Berndt, D.~J. and J.~Clifford (1994).
\newblock Using dynamic time warping to find patterns in time series.
\newblock In {\em KDD workshop}, Volume~10, pp.\  359--370. Seattle, WA, USA:.

\bibitem[\protect\citeauthoryear{Blum and Susarla}{Blum and
  Susarla}{1977}]{blum1977estimation}
Blum, J. and V.~Susarla (1977).
\newblock Estimation of a mixing distribution function.
\newblock {\em The Annals of Probability\/}, 200--209.

\bibitem[\protect\citeauthoryear{Celeux and Soromenho}{Celeux and
  Soromenho}{1996}]{celeux1996entropy}
Celeux, G. and G.~Soromenho (1996).
\newblock An entropy criterion for assessing the number of clusters in a
  mixture model.
\newblock {\em Journal of classification\/}~{\em 13\/}(2), 195--212.

\bibitem[\protect\citeauthoryear{Chandra}{Chandra}{1977}]{chandra1977mixtures}
Chandra, S. (1977).
\newblock On the mixtures of probability distributions.
\newblock {\em Scandinavian Journal of Statistics\/}, 105--112.

\bibitem[\protect\citeauthoryear{Coffey, Hinde, and Holian}{Coffey
  et~al.}{2014}]{coffey2014clustering}
Coffey, N., J.~Hinde, and E.~Holian (2014).
\newblock Clustering longitudinal profiles using p-splines and mixed effects
  models applied to time-course gene expression data.
\newblock {\em Computational Statistics \& Data Analysis\/}~{\em 71}, 14--29.

\bibitem[\protect\citeauthoryear{De~la Cruz-Mes\'ia, Quintana, and
  Marshall}{De~la Cruz-Mes\'ia et~al.}{2008}]{De08}
De~la Cruz-Mes\'ia, R., F.~A. Quintana, and G.~Marshall (2008).
\newblock Model-based clustering for longitudinal data.
\newblock {\em Computational Statistics \& Data Analysis\/}~{\em 52\/}(3),
  1441--1457.

\bibitem[\protect\citeauthoryear{Den~Teuling, Pauws, and van~den
  Heuvel}{Den~Teuling et~al.}{2020}]{den2020comparison}
Den~Teuling, N., S.~Pauws, and E.~van~den Heuvel (2020).
\newblock A comparison of methods for clustering longitudinal data with slowly
  changing trends.
\newblock {\em Communications in Statistics-Simulation and Computation\/},
  1--28.

\bibitem[\protect\citeauthoryear{Genest and Favre}{Genest and
  Favre}{2007}]{Genest07}
Genest, C. and A.-C. Favre (2007).
\newblock Everything you always wanted to know about copula modeling but were
  afraid to ask.
\newblock {\em Journal of hydrologic engineering\/}~{\em 12\/}(4), 347--368.

\bibitem[\protect\citeauthoryear{Genolini, Alacoque, Sentenac, Arnaud,
  et~al.}{Genolini et~al.}{2015}]{Genolini15}
Genolini, C., X.~Alacoque, M.~Sentenac, C.~Arnaud, et~al. (2015).
\newblock kml and kml3d: R packages to cluster longitudinal data.
\newblock {\em Journal of Statistical Software\/}~{\em 65\/}(4), 1--34.

\bibitem[\protect\citeauthoryear{Giacofci, Lambert-Lacroix, Marot, and
  Picard}{Giacofci et~al.}{2013}]{giacofci2013wavelet}
Giacofci, M., S.~Lambert-Lacroix, G.~Marot, and F.~Picard (2013).
\newblock Wavelet-based clustering for mixed-effects functional models in high
  dimension.
\newblock {\em Biometrics\/}~{\em 69\/}(1), 31--40.

\bibitem[\protect\citeauthoryear{Gibbons, Hedeker, and DuToit}{Gibbons
  et~al.}{2010}]{gibbons2010advances}
Gibbons, R.~D., D.~Hedeker, and S.~DuToit (2010).
\newblock Advances in analysis of longitudinal data.
\newblock {\em Annual review of clinical psychology\/}~{\em 6}, 79--107.

\bibitem[\protect\citeauthoryear{Gray}{Gray}{2006}]{gray2006toeplitz}
Gray, R.~M. (2006).
\newblock Toeplitz and circulant matrices: A review.

\bibitem[\protect\citeauthoryear{Huang, Lu, Chen, Liang, and Zangmeister}{Huang
  et~al.}{2018}]{Huang18}
Huang, Y., X.~Lu, J.~Chen, J.~Liang, and M.~Zangmeister (2018).
\newblock Joint model-based clustering of nonlinear longitudinal trajectories
  and associated time-to-event data analysis, linked by latent class
  membership: with application to aids clinical studies.
\newblock {\em Lifetime data analysis\/}~{\em 24\/}(4), 699--718.

\bibitem[\protect\citeauthoryear{Hubert and Arabie}{Hubert and
  Arabie}{1985}]{hubert85}
Hubert, L. and P.~Arabie (1985).
\newblock Comparing partitions.
\newblock {\em Journal of Classification\/}~{\em 2}, 193--218.

\bibitem[\protect\citeauthoryear{Joe}{Joe}{2005}]{Joe05}
Joe, H. (2005).
\newblock Asymptotic efficiency of the two-stage estimation method for
  copula-based models.
\newblock {\em Journal of multivariate Analysis\/}~{\em 94\/}(2), 401--419.

\bibitem[\protect\citeauthoryear{Jones, Nagin, and Roeder}{Jones
  et~al.}{2001}]{Jones01}
Jones, B.~L., D.~S. Nagin, and K.~Roeder (2001).
\newblock A sas procedure based on mixture models for estimating developmental
  trajectories.
\newblock {\em Sociological methods \& research\/}~{\em 29\/}(3), 374--393.

\bibitem[\protect\citeauthoryear{Kayano, Dozono, and Konishi}{Kayano
  et~al.}{2010}]{kayano2010functional}
Kayano, M., K.~Dozono, and S.~Konishi (2010).
\newblock Functional cluster analysis via orthonormalized gaussian basis
  expansions and its application.
\newblock {\em Journal of classification\/}~{\em 27}, 211--230.

\bibitem[\protect\citeauthoryear{Levine, Hunter, and Chauveau}{Levine
  et~al.}{2011}]{levine2011maximum}
Levine, M., D.~R. Hunter, and D.~Chauveau (2011).
\newblock Maximum smoothed likelihood for multivariate mixtures.
\newblock {\em Biometrika\/}~{\em 98\/}(2), 403--416.

\bibitem[\protect\citeauthoryear{McCloud and Parmeter}{McCloud and
  Parmeter}{2020}]{mccloud2020determining}
McCloud, N. and C.~F. Parmeter (2020).
\newblock Determining the number of effective parameters in kernel density
  estimation.
\newblock {\em Computational statistics \& data analysis\/}~{\em 143}, 106843.

\bibitem[\protect\citeauthoryear{McNicholas and Murphy}{McNicholas and
  Murphy}{2010}]{McNicholas10}
McNicholas, P.~D. and T.~B. Murphy (2010).
\newblock Model-based clustering of longitudinal data.
\newblock {\em Canadian Journal of Statistics\/}~{\em 38\/}(1), 153--168.

\bibitem[\protect\citeauthoryear{{R Core Team}}{{R Core Team}}{2023}]{R20}
{R Core Team} (2023).
\newblock {\em R: A Language and Environment for Statistical Computing}.
\newblock Vienna, Austria: R Foundation for Statistical Computing.

\bibitem[\protect\citeauthoryear{Schwarz}{Schwarz}{1978}]{schwarz78}
Schwarz, G. (1978).
\newblock Estimating the dimension of a model.
\newblock {\em Annals of Statistics\/}~{\em 6}, 461--464.

\bibitem[\protect\citeauthoryear{Serban and Wasserman}{Serban and
  Wasserman}{2005}]{serban2005cats}
Serban, N. and L.~Wasserman (2005).
\newblock Cats: clustering after transformation and smoothing.
\newblock {\em Journal of the American Statistical Association\/}~{\em
  100\/}(471), 990--999.

\bibitem[\protect\citeauthoryear{Sklar}{Sklar}{1959}]{sklar59}
Sklar, M. (1959).
\newblock Fonctions de r\'epartition \`a n dimensions et leurs marges.
\newblock {\em Publ. inst. statist. univ. Paris\/}~{\em 8}, 229--231.

\bibitem[\protect\citeauthoryear{Teicher}{Teicher}{1963}]{teicher1963identifiability}
Teicher, H. (1963).
\newblock Identifiability of finite mixtures.
\newblock {\em The annals of Mathematical statistics\/}, 1265--1269.

\bibitem[\protect\citeauthoryear{van~der Nest, Passos, Candel, and van
  Breukelen}{van~der Nest et~al.}{2020}]{van2020overview}
van~der Nest, G., V.~L. Passos, M.~J. Candel, and G.~J. van Breukelen (2020).
\newblock An overview of mixture modelling for latent evolutions in
  longitudinal data: Modelling approaches, fit statistics and software.
\newblock {\em Advances in Life Course Research\/}~{\em 43}, 100323.

\bibitem[\protect\citeauthoryear{Wang, Chiou, and M{\"u}ller}{Wang
  et~al.}{2016}]{wang2016functional}
Wang, J.-L., J.-M. Chiou, and H.-G. M{\"u}ller (2016).
\newblock Functional data analysis.
\newblock {\em Annual Review of Statistics and its application\/}~{\em 3},
  257--295.

\end{thebibliography}
}

\end{document}